\documentclass[prl,aps,twocolumn,superscriptaddress,longbibliography]{revtex4-2}

\usepackage[dvips]{graphicx}
\usepackage{amssymb,amsfonts,amsmath}
\usepackage{color}
\usepackage{ulem}
\usepackage{siunitx}
\usepackage[hidelinks]{hyperref}

\newcommand{\Jdot}{\dot{\textbf{J}}}
\newcommand{\vdot}{\dot{\textbf{v}}}
\newcommand{\J}{\textbf{J}}
\newcommand{\vel}{\textbf{v}}
\newcommand{\acc}{\textbf{a}}
\newcommand{\fext}{\textbf{f}_\text{ext}}
\newcommand{\fvis}{\textbf{f}_\text{vis}}
\newcommand{\fb}{\textbf{f}_{b}}
\newcommand{\fs}{\textbf{f}_{s}}

\newcommand{\cvs}{c^\vel_s}
\newcommand{\cas}{c^\acc_s}
\newcommand{\cvb}{c^\vel_b}
\newcommand{\cab}{c^\acc_b}
\newcommand{\tvs}{\tau^\vel_s}
\newcommand{\tas}{\tau^\acc_s}
\newcommand{\tvb}{\tau^\vel_b}
\newcommand{\tab}{\tau^\acc_b}

\newcommand{\ac}{\textbf{a}}

\newcommand{\rv}{\textbf{r}}

\newcommand{\ex}{\hat{\textbf{e}}_x}
\newcommand{\ey}{\hat{\textbf{e}}_y}

\begin{document}

\title{
       Shear and bulk acceleration viscosities in simple fluids}

\author{Johannes Renner}
\affiliation{Theoretische Physik II, Physikalisches Institut,
  Universit{\"a}t Bayreuth, D-95440 Bayreuth, Germany}
\author{Matthias Schmidt}
\affiliation{Theoretische Physik II, Physikalisches Institut,
  Universit{\"a}t Bayreuth, D-95440 Bayreuth, Germany}
 \author{Daniel de las Heras}
\email{delasheras.daniel@gmail.com}
\homepage{www.danieldelasheras.com}
\affiliation{Theoretische Physik II, Physikalisches Institut,
  Universit{\"a}t Bayreuth, D-95440 Bayreuth, Germany}

\date{\today}

\begin{abstract}
	Inhomogeneities in the velocity field of a moving fluid are dampened by the inherent viscous behaviour of the system.
	Both bulk and shear effects, related to the divergence and the curl of the velocity field, are relevant.
	On molecular time scales, beyond the Navier-Stokes description,	memory plays an important role.
	We demonstrate here on the basis of molecular and overdamped Brownian dynamics many-body simulations that analogous viscous effects act on the acceleration field.
	This acceleration viscous behaviour is associated with the divergence and the curl of the acceleration field and it can be quantitatively described using simple exponentially decaying memory kernels.
	The simultaneous use of velocity and acceleration fields enables the description of fast dynamics on molecular scales. 
\end{abstract}

\maketitle
The viscous force determines the resistance of a moving fluid to change the magnitude and the direction of the flow.
Such viscous response, originated by the interparticle interactions, is relevant in e.g. lubrication~\cite{Bhushan1995}, protein dynamics in biological solvents~\cite{doi:10.1126/science.1615323,https://doi.org/10.1002/jcc.10297}, viscotaxis~\cite{PhysRevLett.120.208002,PhysRevLett.123.158006}, magnetic~\cite{PhysRevLett.75.2128} and quantum~\cite{PhysRevLett.75.697} fluids, lava flows~\cite{doi:10.1146/annurev.fluid.32.1.477}, cardiovascular events~\cite{blood,Kwaan21.11.2003}, food manufacturing~\cite{TABILOMUNIZAGA2005147}, and cosmological models~\cite{Maartens_1995,Gagnon_2011}. Viscous effects are associated with inhomogeneities in the velocity field of the fluid.
The viscous force $\fvis(\rv,t)$ experienced by a particle of a fluid at position $\rv$ and time $t$ contains bulk $\fb(\rv,t)$ and shear $\fs(\rv,t)$ contributions, i.e. $\fvis=\fb+\fs$. These contributions are associated with the divergence $\nabla\cdot\vel$ (bulk) and the curl $\nabla\times\vel$ (shear) of the velocity field $\vel(\rv,t)$, respectively. Specifically, $\fvis$ in the Navier-Stokes~\cite{white2006viscous} equations is
\begin{equation}
	\rho\fvis=\eta_b\nabla\nabla\cdot\vel-\eta_s\nabla\times(\nabla\times\vel),\label{eq:NS}
\end{equation}
where $\rho(\rv,t)$ is the density profile and $\eta_\alpha$ with $\alpha=b,s$ are transport coefficients known as bulk and shear viscosities.

Here, we demonstrate the occurrence in simple fluids of analogue viscous contributions but generated by the divergence and the curl of the acceleration field $\acc(\rv,t)$.
We use custom flow~\cite{de2019custom,customflowMD} to design specific flows (driven by external forces) in which we can unambiguously single out the acceleration contribution of the viscous force.
We consider inhomogeneous and rapidly changing flows. Hence, memory effects and inhomogeneities of the density profile cannot be ignored and need to be included in Eq.~\eqref{eq:NS}.
We propose the following expressions for bulk and shear viscous forces of an inhomogeneous simple fluid 
\begin{align}
	\fb(\rv,t)=\frac{1}{\rho}\int_0^t dt'\left[K_b^{\vel}(t-t')\nabla(\rho\rho'\nabla\cdot\vel')\right.\nonumber\\
	\left.+K_b^{\acc}(t-t')\nabla(\rho\rho'\nabla\cdot\acc')\right],\label{eq:fb}\\
	\fs(\rv,t)=\frac{-1}{\rho}\int_0^t dt'\left[K_s^{\vel}(t-t')\nabla\times(\rho\rho'\nabla\times\vel')\right.\nonumber\\
	\left.+K_s^{\acc}(t-t')\nabla\times(\rho\rho'\nabla\times\acc')\right],\label{eq:fs}
\end{align}
where we leave out the dependence on $\rv$ and $t$, primed quantities are evaluated at $t'$, e.g. $\rho'=\rho(\rv,t')$, and
$K_\alpha^{\boldsymbol\Gamma}$ (with $\alpha={b,s}$ and $\boldsymbol\Gamma={\vel,\acc}$) are exponentially decaying memory kernels
\begin{equation}
	K_\alpha^{\boldsymbol\Gamma}(t-t')=\frac{c_\alpha^{\boldsymbol\Gamma}}{\tau_\alpha^{\boldsymbol\Gamma}}{\rm e}^{-(t-t')/\tau_\alpha^{\boldsymbol\Gamma}},\label{eq:kernel}
\end{equation}
with constant amplitudes $c_\alpha^{\boldsymbol\Gamma}$ and memory times $\tau_\alpha^{\boldsymbol\Gamma}$.
The first terms of Eqs.~\eqref{eq:fb} and~\eqref{eq:fs} are the familiar bulk and shear viscous forces in the Navier-Stokes equations, Eq.~\eqref{eq:NS}, for flows with inhomogeneous density profiles and with the addition of a memory kernel.
The second terms have identical structure but replacing $\vel$ by $\acc$ and represent therefore a viscous response generated by an inhomogeneous acceleration field.
The viscous force in Eq.~\eqref{eq:NS} with viscosities $\eta_\alpha=c^{\vel}_\alpha\rho^2$ follows from 
the velocity contributions of Eqs.~\eqref{eq:fb} and~\eqref{eq:fs} by ignoring the effect of both memory and an inhomogeneous density profile.
Our specific form for $\fvis$ arises in power functional theory~\cite{PowerF,PFTMD,PRLnablaV} by retrieving the first terms of an expansion in acceleration gradients, see additional details in the Supplemental Material (SM)~\cite{Supp}.

\begin{figure*}
  \includegraphics[width=1.0\textwidth]{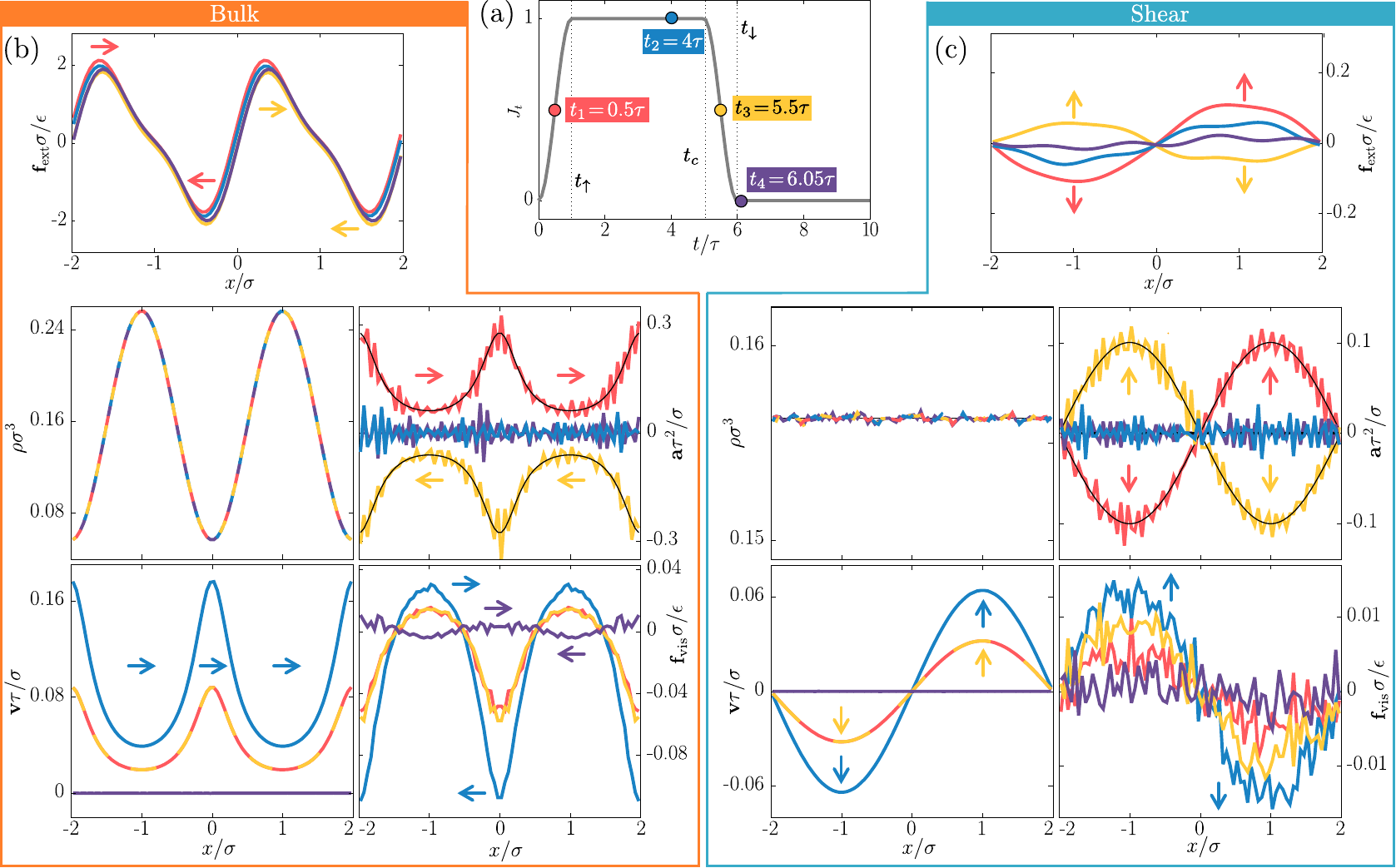}
	\caption{(a) Temporal part of the current $J_t$ vs time $t$ common to the bulk (b) and shear (c) flows. Four times $t_i$ with $i=1,2,3,4$ are highlighted with colored circles.
	The vertical dotted lines indicate the times $t_\uparrow,t_c$, and $t_\downarrow$.
 	Panels (b) and (c) show the external force $\fext$, density $\rho$, velocity $\vel$, acceleration $\acc$, and viscous force $\fvis$ profiles as a function of $x$ for the bulk and shear flows, respectively.
	To improve the visualization, the external force has been smoothed by eliminating high-frequency Fourier modes (see details and raw data in SM~\cite{Supp}).
	The thin black solid lines are the target fields that coincide (up to numerical accuracy) with the sampled fields.
	The color of the profiles indicate the time $t_1=0.5\tau$ (red), $t_2=4\tau$ (blue), $t_3=5.5\tau$ (yellow), and $t_4=6.05\tau$ (purple), as indicated in (a). 
	The arrows indicate the direction of the vector field at specific locations (arrow position) and times (arrow color).
	}
    \label{fig1}
\end{figure*}

To demonstrate the occurrence of viscous effects associated with the acceleration field, we need to disentangle the velocity and the acceleration contributions from the total viscous force.
This requires a complete control over the characteristics of the flow, which we achieve using custom flow~\cite{de2019custom,customflowMD}.
Custom flow uses particle-based simulations to find numerically the spatially and temporally resolved external field required to generate the desired dynamics of a many-body system.
The one-body density $\rho(\rv,t)$ and current $\J(\rv,t)=\rho(\rv,t)\vel(\rv,t)$ profiles serve as input target fields, while the external field $\fext(\rv,t)$ that generates these targets is the output of the method.
At each time, $\fext(\rv,t)$ is constructed iteratively. At iteration $k+1$ we add to the external force of the previous iteration $k$ a term proportional to the difference between target ($\J$) and sampled ($\J^{(k)}$) currents, i.e. $\fext^{(k+1)}=\fext^{(k)}+\alpha_0(\J-\J^{(k)})$. Here, the parameter $\alpha_0(\rv,t)>0$ is chosen to ensure that the difference between target and sampled current fields progressively shrinks.
Details about custom flow are provided in Refs.~\cite{de2019custom,customflowMD} and in SM~\cite{Supp}.
Custom flow is essential here to tailor the dynamics of the system such that the viscous force can be (i) easily measured and (ii) unambiguously split into velocity and acceleration contributions.
We use molecular dynamics (MD) simulations to study a three-dimensional system of particles of mass $m$ interacting via the short-ranged and purely repulsive WCA pair potential~\cite{WCA-potential} with length and energy parameters $\sigma$ and $\epsilon$, respectively.
We work in units of $\sigma$, $\epsilon$, and $m$. Hence, the unit of time is $\tau=\sqrt{m\sigma^2/\epsilon}$.
We consider two different flows that represent pure bulk (compressible) and shear situations.
In both flows the one-body current $\J$ factorizes into a (vectorial) spatial part $\J_\rv$ and a (scalar) temporal part $J_t$, i.e.,\ $\J(\rv,t)=J_t(t)\J_\rv(\rv)$.

The temporal part is common to both flows, see Fig.~\ref{fig1}(a) and SM~\cite{Supp} for the mathematical expression.
The current increases from the initial time until $t_\uparrow=1\tau$, then remains constant (quasi-steady-state) until $t_c=5\tau$, decreases until it vanishes at $t_\downarrow=6\tau$, and it stays zero afterwards. This setup helps to disentangle the velocity and the acceleration contributions from $\fvis$ since $\vel$ and $\acc$ are parallel to each other during the increase of $\J$ but they are antiparallel during the decrease of $\J$. Both $\vel$ and $\acc$ stay unchanged during the quasi-steady-state and during the final evolution towards equilibrium which is useful to characterize memory effects.

Both flows are designed to have a stationary one-body density during the whole time evolution, i.e.\ $\dot\rho(\rv,t)=0$, where the overdot denotes a time derivative.
This simplifies the data analysis since as a direct consequence the viscous forces in Eqs.~\eqref{eq:fb} and~\eqref{eq:fs} also factorize into spatial and temporal terms~\cite{Supp}
\begin{equation}
	\textbf{f}_\alpha(\rv,t)=C_\alpha(t)\textbf{f}_{\rv,\alpha}(\rv),\;\;\alpha={b,s}.\label{eq:split}
\end{equation}

\noindent{\bf Bulk flow.}
Here, by construction $\nabla\times\vel=0$ and $\nabla\times\acc=0$ but $\nabla\cdot\vel\neq0$ and $\nabla\cdot\acc\neq0$.
Hence, only bulk effects contribute to the viscous force, i.e. $\fvis=\fb$.
We take the one-body density to be inhomogeneous but only along the $x$-direction.
The one-body current has only an $x$-component which is taken to be constant in space:
\begin{eqnarray}
	\rho(\rv,t)=\rho(x)&=&\rho_0-\rho_1\cos\left({4\pi x}/{L_x}\right),\\
	\J(\rv,t)=\J(t)&=&J_0J_t(t)\ex,
\end{eqnarray}
with average density $\rho_0\sigma^3=0.15625$, amplitude $\rho_1\sigma^3=0.1$, side length of the simulation box $L_x/\sigma=4$, and maximum value of the current $J_0\tau\sigma^2=0.01$.
Both the velocity $\vel=\J/\rho$ and the acceleration $\acc=\vdot=\Jdot/\rho$ (where the second equality holds here since $\dot\rho=0$) are inhomogeneous in space even though the current is homogeneous.

The external force that produces this bulk flow together with density, velocity, and acceleration profiles sampled in MD are shown in Fig.~\ref{fig1}(b) for four selected times.
The viscous force $\fvis$ [also shown in Fig.~\ref{fig1}(b)] is the part of the internal force that changes sign under flow reversal~\cite{de2020flow,Supp}.
The four times selected in Fig.~\ref{fig1} represent the different regimes of the time evolution imposed by $J_t$, see Fig.~\ref{fig1}(a).
At $t_1=0.5\tau$, i.e.\ $t_1<t_\uparrow$, the current increases and both $\vel$ and $\acc$ point in the same direction.
At $t_2=4\tau$, i.e.\ $t_\uparrow<t_2<t_c$, the system is in a quasi-steady-state with negligible memory effects (we know this by monitoring the viscous force which does not change with time).
The acceleration vanishes everywhere and the velocity profile remains unchanged in this time interval. 
At $t_3=5.5\tau$, i.e.~$t_c<t_3<t_\downarrow$ the current decreases.
The velocity and the acceleration profiles have opposite sign everywhere.
Finally, at $t_4=6.05\tau$, i.e.~$t_4>t_\downarrow$, both $\vel$ and $\acc$ vanish everywhere.
However, due to memory effects the system has not reached equilibrium yet;
there is, for example, a viscous force generated by the history of $\vel$ and $\acc$.

A visual inspection of the viscous force $\fvis$,
Fig.~\ref{fig1}(b), reveals two strong indications that the acceleration
profile contributes to the viscosity. First, at $t_4$ the viscous force
has everywhere the opposite sign than at the previous times.
Hence, the history of the acceleration profile must be
dominating the viscosity since the velocity profile does not
change its sign during the whole time evolution.
Only $\acc$ changes sign during the decrease of the current
[compare the acceleration profiles at times $t_1$ and
$t_3$ in Fig.~\ref{fig1}(b)].
Second, the profiles $\fvis$ at times $t_1$ and $t_3$ are similar.
At these two times the velocity profiles are identical by construction,
see Figs.~\ref{fig1}(a,b). However, $\acc$ and the history
of both $\vel$ and $\acc$ are different.
Since the viscosity at a given time depends on the history of the system,
the contribution to the viscosity due to the acceleration must be cancelling the 
contribution due to the history of the velocity profile. Otherwise, the viscous force
at these times would differ.

\begin{figure}
 \includegraphics[width=1.0\columnwidth]{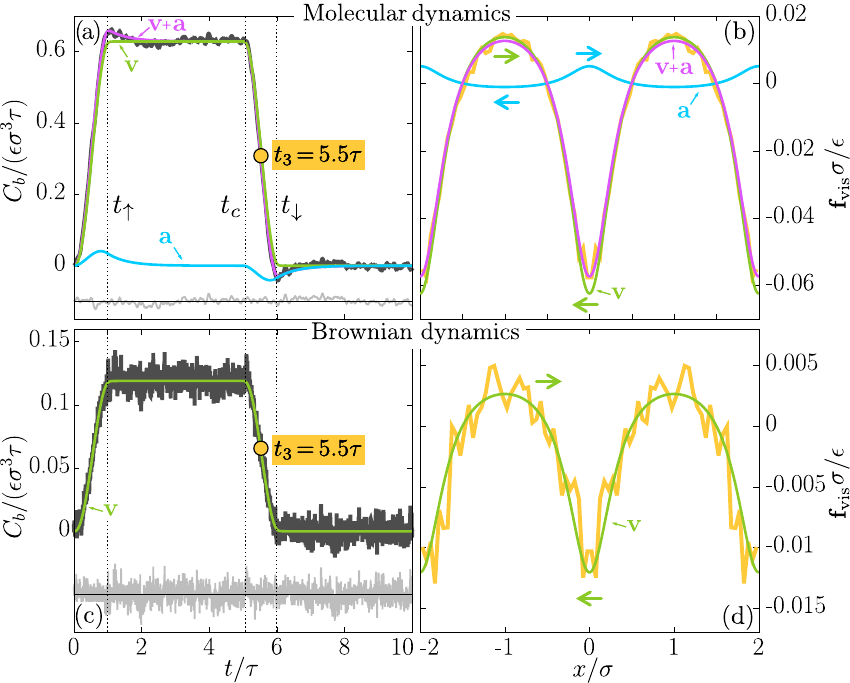}
	\caption{
	(a) Temporal dependency of the bulk viscous force $C_b$ as a function of time $t$ in molecular dynamics simulations (thick black line)
	and theoretically (violet) for the bulk flow. 
	The vertical dotted lines indicate the times $t_\uparrow, t_c$, and $t_\downarrow$.
	The time $t_3=5.5\tau$ is highlighted with a yellow circle.
	The light grey line fluctuating around the horizontal line is the difference between simulation (thick black) and theory (violet).
	(b) Bulk viscous force $\fvis$ as a function of $x$ at time $t_3=5.5\tau$ according to MD (yellow) and theory (violet).
	The force points along the $x$-axis.
	The colored arrows indicate the direction of the corresponding force at selected positions.
	The contributions of the velocity (green) and of the acceleration (blue) to the total signal (violet) are also shown in (a) and (b).
	The bottom panels (c) and (d) show the same data as the top panels but using overdamped Brownian dynamics instead of MD.
	In BD only the velocity field contributes to the viscosity.
	}
	\label{fig2}
\end{figure}

The temporal part $C_b(t)$ for the bulk flow, see Eq.~\eqref{eq:split}, can be understood as the variation of the strength of the viscous force over time. Results are shown in Fig.~\ref{fig2}(a).
Clearly, $C_b$ achieves larger values than at the quasi steady state for times around $t_\uparrow$, and 
smaller (negative) values than in equilibrium ($C_b=0$) for times around $t_\downarrow$.
The acceleration is responsible for the overshoot and the undershoot around the times $t_\uparrow$ and $t_\downarrow$ because $\acc$ is the only field that flips its sign during the increase and during the decrease of the current.
Note that if $\acc$ does not contribute to the bulk viscous force, then the negative values of $C_b$ would indicate an unphysical negative viscosity.

We next compare the MD data to our expression for the viscous force $\fb$, Eq.~\eqref{eq:fb}, to obtain the kernel parameters, see SM~\cite{Supp} for details.
The amplitudes are $\cvb/(\epsilon\sigma^3\tau)=0.63$, $\cab/(\epsilon\sigma^3\tau^2)=0.044,$ and the memory times are $\tvb/\tau=0.043$, $\tab/\tau=0.56$.
The partial contributions of the velocity and the acceleration fields to $C_b$ and $\fvis$ are shown in Fig.~\ref{fig2}(a) and ~\ref{fig2}(b), respectively.
The sum of both contributions agrees quantitatively with the MD data.

To assure that the overshoot and the undershoot in $C_b$ are indeed due to the acceleration field, we performed overdamped Brownian dynamics (BD) simulations for exactly the same flow (using BD custom flow~\cite{de2019custom,Supp} and the usual assumption
that the random force does not depend on the external force~\cite{Kubo_1966}).
Since the system is overdamped, the acceleration does not play any role and indeed there is no overshoot or undershoot in $C_b$, Fig.~\ref{fig2}(c).
Both $C_b$ and $\fvis$ are well reproduced theoretically using only the velocity field, Figs.~\ref{fig2}(c,d), with kernel parameters $\cvb/(\epsilon\sigma^3\tau)=0.117$ and $\tau^\vel_b/\tau=0.041$.

\noindent{\bf Shear flow.} We next consider a flow in which $\nabla\cdot\vel=0$ and $\nabla\cdot\acc=0$ but $\nabla\times\vel\neq0$ and $\nabla\times\acc\neq0$.
Hence, only shear effects contribute to the viscous force, i.e. $\fvis=\fs$.
Using custom flow we set the density profile to be homogeneous and the current to be a shear wave pointing in $y$-direction with modulation along the $x$-direction:
\begin{eqnarray}
	\rho(\rv,t)&=&\rho_0,\\
	\J(\rv,t)&=&\J(x,t)=J_0\sin\left({2\pi x}/{L_x}\right)J_t(t)\ey.\label{eq:shearflow}
\end{eqnarray}
with $\rho_0\sigma^3=0.15625$, $L_x/\sigma=4$, $J_0\tau\sigma^2=0.01$.

\begin{figure}
 \includegraphics[width=1.0\columnwidth]{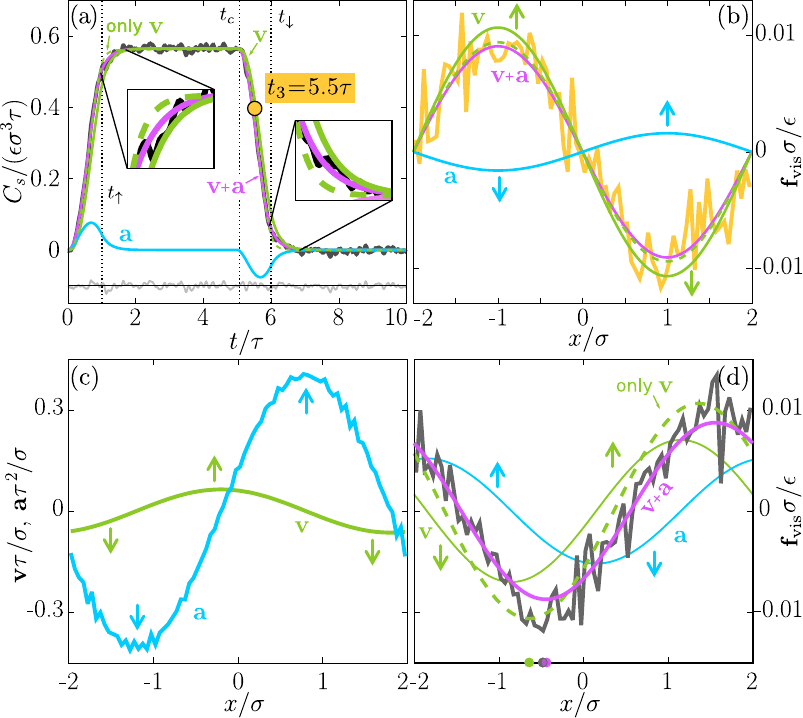}
	\caption{(a) Temporal dependency of the shear viscous force $C_s$ as a function of time $t$ in MD simulations (thick black line)
	and theoretically (violet) for the shear flow. 
	The light grey line fluctuating around the horizontal line is the difference between simulation (thick black) and theory (violet).
	(b) Shear viscous force $\fvis$ as a function of $x$ at time $t_3=5.5\tau$ according to MD (yellow) and theory (violet).
	The force points along the $y-$axis.
	(c) Illustrative velocity (green) and acceleration (blue) profiles vs $x$ for the traveling shear wave ($t=2.7\tau$). Note that $\acc$ and $\vel$ are not in phase.
	(d) Viscous force vs $x$ for the traveling shear wave according to MD (thick black) and theory (violet) ($t=2.7\tau$).
	The colored arrows indicate the direction of the corresponding field at the selected positions.
	The theoretical contributions of $\vel$ (green) and $\acc$ (blue) to the total signal (violet) are also shown in panels (a), (b), and (d) together with the theoretical predictions using only the velocity field (dashed green line). The colored circles over the $x$-axis in (d) indicate the position of the minimum of $\fvis$ according to MD (grey), and theory using both contributions (violet) or only the velocity contribution (green).
	}
	\label{fig3}
\end{figure}

Figure~\ref{fig1}(c) shows the external force required to produce the flow along with results for $\rho,\vel,\ac$, and $\fvis$ at the same four different times as in the previous flow.
A visual inspection of the data does not reveal the acceleration contribution since: (i) for times $t_1=0.5\tau$ and $t_3=5.5\tau$ the curves are different (suggesting either a large memory time of the velocity contribution or a strong effect of the acceleration), and (ii) $\fvis$ does not flip sign after the one-body current vanishes.
Also, in contrast to the bulk flow, no apparent over- or undershoot is present in $C_s(t)$, i.e. the temporal part of $\fvis$, see Fig.~\ref{fig3}(a) and Eq.~\eqref{eq:split}.
For the shear flow we find that the amplitudes $\cvs/(\epsilon\sigma^3\tau)=0.56$, $\cas/(\epsilon\sigma^3\tau^2)=0.059$ and the memory times $\tvs=0.24\tau$, $\tas=0.23\tau$ yield quantitative agreement between simulation data and our theory for both the temporal, Fig.~\ref{fig3}(a), and the spatial dependence of $\fvis$, Fig.~\ref{fig3}(b). In contrast to the bulk flow, the memory times of $\acc$ and $\vel$ are now comparable, which partially hides the effect of the acceleration.
To demonstrate the importance of $\acc$ we use only the velocity contribution and obtain $\cvs/(\epsilon\sigma^3\tau)=0.56$ and  $\tvs/\tau=0.13$ as the optimal kernel parameters.
The resulting curve for $C_s$, see Fig.~\ref{fig3}(a), deviates from the MD data around the times $t_\uparrow$ (curve above MD data) and $t_\downarrow$ (curve below MD data).
This indicates that $\acc$ indeed contributes since its sign change around these times can correct these deviations.

To further ascertain the reality of the acceleration contribution, we use the obtained parameters for the amplitudes and the memory times to describe a variation of the flow.
Instead of decreasing the one-body current after $t_c$, we keep the amplitude of the current unchanged and let the shear wave travel in the positive $x$-direction.
Specifically, after time $t=2\tau>t_\uparrow$ we replace the $x$ coordinate in Eq.~\eqref{eq:shearflow} by $x-v_st$ with constant velocity $v_s=4\tau/\sigma$.
Hence, the acceleration field is shifted by $\pi/2$ with respect to the velocity field, see Fig.~\ref{fig3}(c).
The phase difference between $\vel$ and $\acc$ has an effect on the viscous force, see Fig.~\ref{fig3}(d).
Using the kernel parameters for the previous flow and both the velocity and the acceleration contributions we reproduce the simulation data.
In contrast, using the parameters obtained only with the velocity contribution results in a clear phase shift compared to the MD data.
See SM~\cite{Supp} for more details.

Our results demonstrate the existence of shear and bulk acceleration viscous forces generated by inhomogeneities of the acceleration field.
These forces act in addition to the usual viscous response associated with the velocity field.
In our examples the contribution of the acceleration to the viscous force is quantitatively significant.
Acceleration viscous forces might be also relevant in flows with rapid temporal changes of the velocity field such as in shock waves~\cite{holian1980shock,holian1988modeling,PhysRevLett.69.269,PhysRevLett.83.1175,PhysRevLett.118.025001}, turbulent flows~\cite{RevModPhys.73.913,Bentkamp2019,PhysRevLett.127.074501} including atmospheric and oceanic flows~\cite{vallis2017atmospheric}, inertial microfluidics~\cite{C3LC51403J,B912547G,C5LC01159K}, the description of flows at the nanoscale~\cite{doi:10.1063/1.1559936,Straube2020,doi:10.1146/annurev-fluid-071320-095958}, mudflows~\cite{coussot2017mudflow}, single-bubble sonoluminescence~\cite{RevModPhys.74.425,PhysRevLett.96.114301}, and viscous cosmological models~\cite{PhysRevD.48.1597,Maartens_1995}.

We did not use a thermostat due to the low heat production in both flows (the temperature increase was less than $2\%$ from the initial to the final state~\cite{Supp}).
However, custom flow can be used with thermostats~\cite{customflowMD}, and it would be interesting to compare the effect of the acceleration viscosities in thermalized and 
non-thermalized flows.

We use here a rather simple kernel as compared to other approaches~\cite{jung2017iterative,meyer2020non,PhysRevLett.116.147804,PhysRevX.7.041065}.
The use of simple memory kernels that decay exponentially in time is only possible because we use all physically relevant variables, i.e. both $\vel$ and $\acc$. 
Since $\acc$ and $\vel$ are related to each other, it should be possible to describe $\fvis$ using only $\vel$ or $\acc$ together with a complicated kernel.
Such kernel would be tailored to the specific flow instead of being general to every situation.
For example, it might be possible to describe the viscous force of the bulk flow using only $\vel$ and a complex memory kernel with a negative tail.

\begin{acknowledgments}
This work is supported by the German Research Foundation (DFG) via
project number 447925252.
\end{acknowledgments}


\begin{thebibliography}{46}%
\makeatletter
\providecommand \@ifxundefined [1]{%
 \@ifx{#1\undefined}
}%
\providecommand \@ifnum [1]{%
 \ifnum #1\expandafter \@firstoftwo
 \else \expandafter \@secondoftwo
 \fi
}%
\providecommand \@ifx [1]{%
 \ifx #1\expandafter \@firstoftwo
 \else \expandafter \@secondoftwo
 \fi
}%
\providecommand \natexlab [1]{#1}%
\providecommand \enquote  [1]{``#1''}%
\providecommand \bibnamefont  [1]{#1}%
\providecommand \bibfnamefont [1]{#1}%
\providecommand \citenamefont [1]{#1}%
\providecommand \href@noop [0]{\@secondoftwo}%
\providecommand \href [0]{\begingroup \@sanitize@url \@href}%
\providecommand \@href[1]{\@@startlink{#1}\@@href}%
\providecommand \@@href[1]{\endgroup#1\@@endlink}%
\providecommand \@sanitize@url [0]{\catcode `\\12\catcode `\$12\catcode
  `\&12\catcode `\#12\catcode `\^12\catcode `\_12\catcode `\%12\relax}%
\providecommand \@@startlink[1]{}%
\providecommand \@@endlink[0]{}%
\providecommand \url  [0]{\begingroup\@sanitize@url \@url }%
\providecommand \@url [1]{\endgroup\@href {#1}{\urlprefix }}%
\providecommand \urlprefix  [0]{URL }%
\providecommand \Eprint [0]{\href }%
\providecommand \doibase [0]{https://doi.org/}%
\providecommand \selectlanguage [0]{\@gobble}%
\providecommand \bibinfo  [0]{\@secondoftwo}%
\providecommand \bibfield  [0]{\@secondoftwo}%
\providecommand \translation [1]{[#1]}%
\providecommand \BibitemOpen [0]{}%
\providecommand \bibitemStop [0]{}%
\providecommand \bibitemNoStop [0]{.\EOS\space}%
\providecommand \EOS [0]{\spacefactor3000\relax}%
\providecommand \BibitemShut  [1]{\csname bibitem#1\endcsname}%
\let\auto@bib@innerbib\@empty
\bibitem [{\citenamefont {Bhushan}\ \emph {et~al.}(1995)\citenamefont
  {Bhushan}, \citenamefont {Israelachvili},\ and\ \citenamefont
  {Landman}}]{Bhushan1995}%
  \BibitemOpen
  \bibfield  {author} {\bibinfo {author} {\bibfnamefont {B.}~\bibnamefont
  {Bhushan}}, \bibinfo {author} {\bibfnamefont {J.~N.}\ \bibnamefont
  {Israelachvili}},\ and\ \bibinfo {author} {\bibfnamefont {U.}~\bibnamefont
  {Landman}},\ }\bibfield  {title} {\bibinfo {title} {Nanotribology: friction,
  wear and lubrication at the atomic scale},\ }\href
  {https://doi.org/10.1038/374607a0} {\bibfield  {journal} {\bibinfo  {journal}
  {Nature}\ }\textbf {\bibinfo {volume} {374}},\ \bibinfo {pages} {607}
  (\bibinfo {year} {1995})}\BibitemShut {NoStop}%
\bibitem [{\citenamefont {Ansari}\ \emph {et~al.}(1992)\citenamefont {Ansari},
  \citenamefont {Jones}, \citenamefont {Henry}, \citenamefont {Hofrichter},\
  and\ \citenamefont {Eaton}}]{doi:10.1126/science.1615323}%
  \BibitemOpen
  \bibfield  {author} {\bibinfo {author} {\bibfnamefont {A.}~\bibnamefont
  {Ansari}}, \bibinfo {author} {\bibfnamefont {C.~M.}\ \bibnamefont {Jones}},
  \bibinfo {author} {\bibfnamefont {E.~R.}\ \bibnamefont {Henry}}, \bibinfo
  {author} {\bibfnamefont {J.}~\bibnamefont {Hofrichter}},\ and\ \bibinfo
  {author} {\bibfnamefont {W.~A.}\ \bibnamefont {Eaton}},\ }\bibfield  {title}
  {\bibinfo {title} {The role of solvent viscosity in the dynamics of protein
  conformational changes},\ }\href {https://doi.org/10.1126/science.1615323}
  {\bibfield  {journal} {\bibinfo  {journal} {Science}\ }\textbf {\bibinfo
  {volume} {256}},\ \bibinfo {pages} {1796} (\bibinfo {year}
  {1992})}\BibitemShut {NoStop}%
\bibitem [{\citenamefont {Zagrovic}\ and\ \citenamefont
  {Pande}(2003)}]{https://doi.org/10.1002/jcc.10297}%
  \BibitemOpen
  \bibfield  {author} {\bibinfo {author} {\bibfnamefont {B.}~\bibnamefont
  {Zagrovic}}\ and\ \bibinfo {author} {\bibfnamefont {V.}~\bibnamefont
  {Pande}},\ }\bibfield  {title} {\bibinfo {title} {Solvent viscosity
  dependence of the folding rate of a small protein: Distributed computing
  study},\ }\href {https://doi.org/https://doi.org/10.1002/jcc.10297}
  {\bibfield  {journal} {\bibinfo  {journal} {J. Comput. Chem.}\ }\textbf
  {\bibinfo {volume} {24}},\ \bibinfo {pages} {1432} (\bibinfo {year}
  {2003})}\BibitemShut {NoStop}%
\bibitem [{\citenamefont {Liebchen}\ \emph {et~al.}(2018)\citenamefont
  {Liebchen}, \citenamefont {Monderkamp}, \citenamefont {ten Hagen},\ and\
  \citenamefont {L\"owen}}]{PhysRevLett.120.208002}%
  \BibitemOpen
  \bibfield  {author} {\bibinfo {author} {\bibfnamefont {B.}~\bibnamefont
  {Liebchen}}, \bibinfo {author} {\bibfnamefont {P.}~\bibnamefont
  {Monderkamp}}, \bibinfo {author} {\bibfnamefont {B.}~\bibnamefont {ten
  Hagen}},\ and\ \bibinfo {author} {\bibfnamefont {H.}~\bibnamefont
  {L\"owen}},\ }\bibfield  {title} {\bibinfo {title} {Viscotaxis: Microswimmer
  navigation in viscosity gradients},\ }\href
  {https://doi.org/10.1103/PhysRevLett.120.208002} {\bibfield  {journal}
  {\bibinfo  {journal} {Phys. Rev. Lett.}\ }\textbf {\bibinfo {volume} {120}},\
  \bibinfo {pages} {208002} (\bibinfo {year} {2018})}\BibitemShut {NoStop}%
\bibitem [{\citenamefont {Datt}\ and\ \citenamefont
  {Elfring}(2019)}]{PhysRevLett.123.158006}%
  \BibitemOpen
  \bibfield  {author} {\bibinfo {author} {\bibfnamefont {C.}~\bibnamefont
  {Datt}}\ and\ \bibinfo {author} {\bibfnamefont {G.~J.}\ \bibnamefont
  {Elfring}},\ }\bibfield  {title} {\bibinfo {title} {Active particles in
  viscosity gradients},\ }\href
  {https://doi.org/10.1103/PhysRevLett.123.158006} {\bibfield  {journal}
  {\bibinfo  {journal} {Phys. Rev. Lett.}\ }\textbf {\bibinfo {volume} {123}},\
  \bibinfo {pages} {158006} (\bibinfo {year} {2019})}\BibitemShut {NoStop}%
\bibitem [{\citenamefont {Bacri}\ \emph {et~al.}(1995)\citenamefont {Bacri},
  \citenamefont {Perzynski}, \citenamefont {Shliomis},\ and\ \citenamefont
  {Burde}}]{PhysRevLett.75.2128}%
  \BibitemOpen
  \bibfield  {author} {\bibinfo {author} {\bibfnamefont {J.-C.}\ \bibnamefont
  {Bacri}}, \bibinfo {author} {\bibfnamefont {R.}~\bibnamefont {Perzynski}},
  \bibinfo {author} {\bibfnamefont {M.~I.}\ \bibnamefont {Shliomis}},\ and\
  \bibinfo {author} {\bibfnamefont {G.~I.}\ \bibnamefont {Burde}},\ }\bibfield
  {title} {\bibinfo {title} {``{N}egative-viscosity'' effect in a magnetic
  fluid},\ }\href {https://doi.org/10.1103/PhysRevLett.75.2128} {\bibfield
  {journal} {\bibinfo  {journal} {Phys. Rev. Lett.}\ }\textbf {\bibinfo
  {volume} {75}},\ \bibinfo {pages} {2128} (\bibinfo {year}
  {1995})}\BibitemShut {NoStop}%
\bibitem [{\citenamefont {Avron}\ \emph {et~al.}(1995)\citenamefont {Avron},
  \citenamefont {Seiler},\ and\ \citenamefont {Zograf}}]{PhysRevLett.75.697}%
  \BibitemOpen
  \bibfield  {author} {\bibinfo {author} {\bibfnamefont {J.~E.}\ \bibnamefont
  {Avron}}, \bibinfo {author} {\bibfnamefont {R.}~\bibnamefont {Seiler}},\ and\
  \bibinfo {author} {\bibfnamefont {P.~G.}\ \bibnamefont {Zograf}},\ }\bibfield
   {title} {\bibinfo {title} {Viscosity of quantum hall fluids},\ }\href
  {https://doi.org/10.1103/PhysRevLett.75.697} {\bibfield  {journal} {\bibinfo
  {journal} {Phys. Rev. Lett.}\ }\textbf {\bibinfo {volume} {75}},\ \bibinfo
  {pages} {697} (\bibinfo {year} {1995})}\BibitemShut {NoStop}%
\bibitem [{\citenamefont
  {Griffiths}(2000)}]{doi:10.1146/annurev.fluid.32.1.477}%
  \BibitemOpen
  \bibfield  {author} {\bibinfo {author} {\bibfnamefont {R.~W.}\ \bibnamefont
  {Griffiths}},\ }\bibfield  {title} {\bibinfo {title} {The dynamics of lava
  flows},\ }\href {https://doi.org/10.1146/annurev.fluid.32.1.477} {\bibfield
  {journal} {\bibinfo  {journal} {Annu. Rev. Fluid Mech.}\ }\textbf {\bibinfo
  {volume} {32}},\ \bibinfo {pages} {477} (\bibinfo {year} {2000})}\BibitemShut
  {NoStop}%
\bibitem [{\citenamefont {Lowe}\ \emph {et~al.}(1997)\citenamefont {Lowe},
  \citenamefont {Lee}, \citenamefont {Rumley}, \citenamefont {Price},\ and\
  \citenamefont {Fowkes}}]{blood}%
  \BibitemOpen
  \bibfield  {author} {\bibinfo {author} {\bibfnamefont {G.~D.~O.}\
  \bibnamefont {Lowe}}, \bibinfo {author} {\bibfnamefont {A.~J.}\ \bibnamefont
  {Lee}}, \bibinfo {author} {\bibfnamefont {A.}~\bibnamefont {Rumley}},
  \bibinfo {author} {\bibfnamefont {J.~F.}\ \bibnamefont {Price}},\ and\
  \bibinfo {author} {\bibfnamefont {F.~G.~R.}\ \bibnamefont {Fowkes}},\
  }\bibfield  {title} {\bibinfo {title} {Blood viscosity and risk of
  cardiovascular events: the {E}dinburgh {A}rtery {S}tudy},\ }\href
  {https://doi.org/https://doi.org/10.1046/j.1365-2141.1997.8532481.x}
  {\bibfield  {journal} {\bibinfo  {journal} {Br. J. Haematol.}\ }\textbf
  {\bibinfo {volume} {96}},\ \bibinfo {pages} {168} (\bibinfo {year}
  {1997})}\BibitemShut {NoStop}%
\bibitem [{\citenamefont {Kwaan}(2003)}]{Kwaan21.11.2003}%
  \BibitemOpen
  \bibfield  {author} {\bibinfo {author} {\bibfnamefont {H.~C.}\ \bibnamefont
  {Kwaan}},\ }\bibfield  {title} {\bibinfo {title} {The {H}yperviscosity
  {S}yndromes},\ }\href {https://doi.org/10.1055/s-2003-44550} {\bibfield
  {journal} {\bibinfo  {journal} {Semin. Thromb. Hemost.}\ }\textbf {\bibinfo
  {volume} {29}},\ \bibinfo {pages} {433} (\bibinfo {year} {2003})}\BibitemShut
  {NoStop}%
\bibitem [{\citenamefont {Tabilo-Munizaga}\ and\ \citenamefont
  {Barbosa-Cánovas}(2005)}]{TABILOMUNIZAGA2005147}%
  \BibitemOpen
  \bibfield  {author} {\bibinfo {author} {\bibfnamefont {G.}~\bibnamefont
  {Tabilo-Munizaga}}\ and\ \bibinfo {author} {\bibfnamefont {G.~V.}\
  \bibnamefont {Barbosa-Cánovas}},\ }\bibfield  {title} {\bibinfo {title}
  {Rheology for the food industry},\ }\href
  {https://doi.org/https://doi.org/10.1016/j.jfoodeng.2004.05.062} {\bibfield
  {journal} {\bibinfo  {journal} {J. Food Eng.}\ }\textbf {\bibinfo {volume}
  {67}},\ \bibinfo {pages} {147} (\bibinfo {year} {2005})}\BibitemShut
  {NoStop}%
\bibitem [{\citenamefont {Maartens}(1995)}]{Maartens_1995}%
  \BibitemOpen
  \bibfield  {author} {\bibinfo {author} {\bibfnamefont {R.}~\bibnamefont
  {Maartens}},\ }\bibfield  {title} {\bibinfo {title} {Dissipative cosmology},\
  }\href {https://doi.org/10.1088/0264-9381/12/6/011} {\bibfield  {journal}
  {\bibinfo  {journal} {Class. Quantum Grav.}\ }\textbf {\bibinfo {volume}
  {12}},\ \bibinfo {pages} {1455} (\bibinfo {year} {1995})}\BibitemShut
  {NoStop}%
\bibitem [{\citenamefont {Gagnon}\ and\ \citenamefont
  {Lesgourgues}(2011)}]{Gagnon_2011}%
  \BibitemOpen
  \bibfield  {author} {\bibinfo {author} {\bibfnamefont {J.-S.}\ \bibnamefont
  {Gagnon}}\ and\ \bibinfo {author} {\bibfnamefont {J.}~\bibnamefont
  {Lesgourgues}},\ }\bibfield  {title} {\bibinfo {title} {Dark goo: bulk
  viscosity as an alternative to dark energy},\ }\href
  {https://doi.org/10.1088/1475-7516/2011/09/026} {\bibfield  {journal}
  {\bibinfo  {journal} {J. Cosmol. Astropart. Phys.}\ }\textbf {\bibinfo
  {volume} {2011}}\bibinfo  {number} { (09)},\ \bibinfo {pages}
  {026}}\BibitemShut {NoStop}%
\bibitem [{\citenamefont {White}\ and\ \citenamefont
  {Majdalani}(2006)}]{white2006viscous}%
  \BibitemOpen
\bibfield  {number} {  }\bibfield  {author} {\bibinfo {author} {\bibfnamefont
  {F.~M.}\ \bibnamefont {White}}\ and\ \bibinfo {author} {\bibfnamefont
  {J.}~\bibnamefont {Majdalani}},\ }\href@noop {} {\emph {\bibinfo {title}
  {Viscous fluid flow}}},\ Vol.~\bibinfo {volume} {3}\ (\bibinfo  {publisher}
  {McGraw-Hill New York},\ \bibinfo {year} {2006})\BibitemShut {NoStop}%
\bibitem [{\citenamefont {de~las Heras}\ \emph {et~al.}(2019)\citenamefont
  {de~las Heras}, \citenamefont {Renner},\ and\ \citenamefont
  {Schmidt}}]{de2019custom}%
  \BibitemOpen
  \bibfield  {author} {\bibinfo {author} {\bibfnamefont {D.}~\bibnamefont
  {de~las Heras}}, \bibinfo {author} {\bibfnamefont {J.}~\bibnamefont
  {Renner}},\ and\ \bibinfo {author} {\bibfnamefont {M.}~\bibnamefont
  {Schmidt}},\ }\bibfield  {title} {\bibinfo {title} {Custom flow in overdamped
  {B}rownian dynamics},\ }\href {https://doi.org/10.1103/PhysRevE.99.023306}
  {\bibfield  {journal} {\bibinfo  {journal} {Phys. Rev. E}\ }\textbf {\bibinfo
  {volume} {99}},\ \bibinfo {pages} {023306} (\bibinfo {year}
  {2019})}\BibitemShut {NoStop}%
\bibitem [{\citenamefont {Renner}\ \emph {et~al.}(2021)\citenamefont {Renner},
  \citenamefont {Schmidt},\ and\ \citenamefont {de~las Heras}}]{customflowMD}%
  \BibitemOpen
  \bibfield  {author} {\bibinfo {author} {\bibfnamefont {J.}~\bibnamefont
  {Renner}}, \bibinfo {author} {\bibfnamefont {M.}~\bibnamefont {Schmidt}},\
  and\ \bibinfo {author} {\bibfnamefont {D.}~\bibnamefont {de~las Heras}},\
  }\bibfield  {title} {\bibinfo {title} {Custom flow in molecular dynamics},\
  }\href {https://doi.org/10.1103/PhysRevResearch.3.013281} {\bibfield
  {journal} {\bibinfo  {journal} {Phys. Rev. Res.}\ }\textbf {\bibinfo {volume}
  {3}},\ \bibinfo {pages} {013281} (\bibinfo {year} {2021})}\BibitemShut
  {NoStop}%
\bibitem [{\citenamefont {Schmidt}\ and\ \citenamefont
  {Brader}(2013)}]{PowerF}%
  \BibitemOpen
  \bibfield  {author} {\bibinfo {author} {\bibfnamefont {M.}~\bibnamefont
  {Schmidt}}\ and\ \bibinfo {author} {\bibfnamefont {J.~M.}\ \bibnamefont
  {Brader}},\ }\bibfield  {title} {\bibinfo {title} {Power functional theory
  for brownian dynamics},\ }\href
  {https://doi.org/http://dx.doi.org/10.1063/1.4807586} {\bibfield  {journal}
  {\bibinfo  {journal} {J. Chem. Phys.}\ }\textbf {\bibinfo {volume} {138}},\
  \bibinfo {pages} {214101} (\bibinfo {year} {2013})}\BibitemShut {NoStop}%
\bibitem [{\citenamefont {Schmidt}(2018)}]{PFTMD}%
  \BibitemOpen
  \bibfield  {author} {\bibinfo {author} {\bibfnamefont {M.}~\bibnamefont
  {Schmidt}},\ }\bibfield  {title} {\bibinfo {title} {Power functional theory
  for {N}ewtonian many-body dynamics},\ }\href
  {https://doi.org/10.1063/1.5008608} {\bibfield  {journal} {\bibinfo
  {journal} {J. Chem. Phys.}\ }\textbf {\bibinfo {volume} {148}},\ \bibinfo
  {pages} {044502} (\bibinfo {year} {2018})}\BibitemShut {NoStop}%
\bibitem [{\citenamefont {de~las Heras}\ and\ \citenamefont
  {Schmidt}(2018)}]{PRLnablaV}%
  \BibitemOpen
  \bibfield  {author} {\bibinfo {author} {\bibfnamefont {D.}~\bibnamefont
  {de~las Heras}}\ and\ \bibinfo {author} {\bibfnamefont {M.}~\bibnamefont
  {Schmidt}},\ }\bibfield  {title} {\bibinfo {title} {Velocity gradient power
  functional for {Brownian} dynamics},\ }\href
  {https://doi.org/10.1103/PhysRevLett.120.028001} {\bibfield  {journal}
  {\bibinfo  {journal} {Phys. Rev. Lett.}\ }\textbf {\bibinfo {volume} {120}},\
  \bibinfo {pages} {028001} (\bibinfo {year} {2018})}\BibitemShut {NoStop}%
\bibitem [{Sup()}]{Supp}%
  \BibitemOpen
  \href@noop {} {\bibinfo  {journal} {See Supplemental Material for details
  about simulations, custom flow, and power functional theory}\ }\BibitemShut
  {NoStop}%
\bibitem [{\citenamefont {Weeks}\ \emph {et~al.}(1971)\citenamefont {Weeks},
  \citenamefont {Chandler},\ and\ \citenamefont {Andersen}}]{WCA-potential}%
  \BibitemOpen
\bibfield  {journal} {  }\bibfield  {author} {\bibinfo {author} {\bibfnamefont
  {J.~D.}\ \bibnamefont {Weeks}}, \bibinfo {author} {\bibfnamefont
  {D.}~\bibnamefont {Chandler}},\ and\ \bibinfo {author} {\bibfnamefont
  {H.~C.}\ \bibnamefont {Andersen}},\ }\bibfield  {title} {\bibinfo {title}
  {Role of repulsive forces in determining the equilibrium structure of simple
  liquids},\ }\href {https://doi.org/10.1063/1.1674820} {\bibfield  {journal}
  {\bibinfo  {journal} {J. Chem. Phys}\ }\textbf {\bibinfo {volume} {54}},\
  \bibinfo {pages} {5237} (\bibinfo {year} {1971})}\BibitemShut {NoStop}%
\bibitem [{\citenamefont {de~las Heras}\ and\ \citenamefont
  {Schmidt}(2020)}]{de2020flow}%
  \BibitemOpen
  \bibfield  {author} {\bibinfo {author} {\bibfnamefont {D.}~\bibnamefont
  {de~las Heras}}\ and\ \bibinfo {author} {\bibfnamefont {M.}~\bibnamefont
  {Schmidt}},\ }\bibfield  {title} {\bibinfo {title} {Flow and structure in
  nonequilibrium brownian many-body systems},\ }\href
  {https://doi.org/10.1103/PhysRevLett.125.018001} {\bibfield  {journal}
  {\bibinfo  {journal} {Phys. Rev. Lett.}\ }\textbf {\bibinfo {volume} {125}},\
  \bibinfo {pages} {018001} (\bibinfo {year} {2020})}\BibitemShut {NoStop}%
\bibitem [{\citenamefont {Kubo}(1966)}]{Kubo_1966}%
  \BibitemOpen
  \bibfield  {author} {\bibinfo {author} {\bibfnamefont {R.}~\bibnamefont
  {Kubo}},\ }\bibfield  {title} {\bibinfo {title} {The fluctuation-dissipation
  theorem},\ }\href {https://doi.org/10.1088/0034-4885/29/1/306} {\bibfield
  {journal} {\bibinfo  {journal} {Rep. Prog. Phys.}\ }\textbf {\bibinfo
  {volume} {29}},\ \bibinfo {pages} {255} (\bibinfo {year} {1966})}\BibitemShut
  {NoStop}%
\bibitem [{\citenamefont {Holian}\ \emph {et~al.}(1980)\citenamefont {Holian},
  \citenamefont {Hoover}, \citenamefont {Moran},\ and\ \citenamefont
  {Straub}}]{holian1980shock}%
  \BibitemOpen
  \bibfield  {author} {\bibinfo {author} {\bibfnamefont {B.~L.}\ \bibnamefont
  {Holian}}, \bibinfo {author} {\bibfnamefont {W.~G.}\ \bibnamefont {Hoover}},
  \bibinfo {author} {\bibfnamefont {B.}~\bibnamefont {Moran}},\ and\ \bibinfo
  {author} {\bibfnamefont {G.~K.}\ \bibnamefont {Straub}},\ }\bibfield  {title}
  {\bibinfo {title} {Shock-wave structure via nonequilibrium molecular dynamics
  and {N}avier-{S}tokes continuum mechanics},\ }\href
  {https://doi.org/10.1103/PhysRevA.22.2798} {\bibfield  {journal} {\bibinfo
  {journal} {Phys. Rev. A}\ }\textbf {\bibinfo {volume} {22}},\ \bibinfo
  {pages} {2798} (\bibinfo {year} {1980})}\BibitemShut {NoStop}%
\bibitem [{\citenamefont {Holian}(1988)}]{holian1988modeling}%
  \BibitemOpen
  \bibfield  {author} {\bibinfo {author} {\bibfnamefont {B.~L.}\ \bibnamefont
  {Holian}},\ }\bibfield  {title} {\bibinfo {title} {Modeling shock-wave
  deformation via molecular dynamics},\ }\href
  {https://doi.org/10.1103/PhysRevA.37.2562} {\bibfield  {journal} {\bibinfo
  {journal} {Phys. Rev. A}\ }\textbf {\bibinfo {volume} {37}},\ \bibinfo
  {pages} {2562} (\bibinfo {year} {1988})}\BibitemShut {NoStop}%
\bibitem [{\citenamefont {Salomons}\ and\ \citenamefont
  {Mareschal}(1992)}]{PhysRevLett.69.269}%
  \BibitemOpen
  \bibfield  {author} {\bibinfo {author} {\bibfnamefont {E.}~\bibnamefont
  {Salomons}}\ and\ \bibinfo {author} {\bibfnamefont {M.}~\bibnamefont
  {Mareschal}},\ }\bibfield  {title} {\bibinfo {title} {Usefulness of the
  {B}urnett description of strong shock waves},\ }\href
  {https://doi.org/10.1103/PhysRevLett.69.269} {\bibfield  {journal} {\bibinfo
  {journal} {Phys. Rev. Lett.}\ }\textbf {\bibinfo {volume} {69}},\ \bibinfo
  {pages} {269} (\bibinfo {year} {1992})}\BibitemShut {NoStop}%
\bibitem [{\citenamefont {Zhakhovski\ifmmode~\breve{\imath}\else \u{\i}\fi{}}\
  \emph {et~al.}(1999)\citenamefont {Zhakhovski\ifmmode~\breve{\imath}\else
  \u{\i}\fi{}}, \citenamefont {Zybin}, \citenamefont {Nishihara},\ and\
  \citenamefont {Anisimov}}]{PhysRevLett.83.1175}%
  \BibitemOpen
  \bibfield  {author} {\bibinfo {author} {\bibfnamefont {V.~V.}\ \bibnamefont
  {Zhakhovski\ifmmode~\breve{\imath}\else \u{\i}\fi{}}}, \bibinfo {author}
  {\bibfnamefont {S.~V.}\ \bibnamefont {Zybin}}, \bibinfo {author}
  {\bibfnamefont {K.}~\bibnamefont {Nishihara}},\ and\ \bibinfo {author}
  {\bibfnamefont {S.~I.}\ \bibnamefont {Anisimov}},\ }\bibfield  {title}
  {\bibinfo {title} {Shock wave structure in {L}ennard-{J}ones crystal via
  molecular dynamics},\ }\href {https://doi.org/10.1103/PhysRevLett.83.1175}
  {\bibfield  {journal} {\bibinfo  {journal} {Phys. Rev. Lett.}\ }\textbf
  {\bibinfo {volume} {83}},\ \bibinfo {pages} {1175} (\bibinfo {year}
  {1999})}\BibitemShut {NoStop}%
\bibitem [{\citenamefont {Marciante}\ and\ \citenamefont
  {Murillo}(2017)}]{PhysRevLett.118.025001}%
  \BibitemOpen
  \bibfield  {author} {\bibinfo {author} {\bibfnamefont {M.}~\bibnamefont
  {Marciante}}\ and\ \bibinfo {author} {\bibfnamefont {M.~S.}\ \bibnamefont
  {Murillo}},\ }\bibfield  {title} {\bibinfo {title} {Thermodynamic and kinetic
  properties of shocks in two-dimensional {Y}ukawa systems},\ }\href
  {https://doi.org/10.1103/PhysRevLett.118.025001} {\bibfield  {journal}
  {\bibinfo  {journal} {Phys. Rev. Lett.}\ }\textbf {\bibinfo {volume} {118}},\
  \bibinfo {pages} {025001} (\bibinfo {year} {2017})}\BibitemShut {NoStop}%
\bibitem [{\citenamefont {Falkovich}\ \emph {et~al.}(2001)\citenamefont
  {Falkovich}, \citenamefont {Gaw\ifmmode~\mbox{\c{e}}\else \c{e}\fi{}dzki},\
  and\ \citenamefont {Vergassola}}]{RevModPhys.73.913}%
  \BibitemOpen
  \bibfield  {author} {\bibinfo {author} {\bibfnamefont {G.}~\bibnamefont
  {Falkovich}}, \bibinfo {author} {\bibfnamefont {K.}~\bibnamefont
  {Gaw\ifmmode~\mbox{\c{e}}\else \c{e}\fi{}dzki}},\ and\ \bibinfo {author}
  {\bibfnamefont {M.}~\bibnamefont {Vergassola}},\ }\bibfield  {title}
  {\bibinfo {title} {Particles and fields in fluid turbulence},\ }\href
  {https://doi.org/10.1103/RevModPhys.73.913} {\bibfield  {journal} {\bibinfo
  {journal} {Rev. Mod. Phys.}\ }\textbf {\bibinfo {volume} {73}},\ \bibinfo
  {pages} {913} (\bibinfo {year} {2001})}\BibitemShut {NoStop}%
\bibitem [{\citenamefont {Bentkamp}\ \emph {et~al.}(2019)\citenamefont
  {Bentkamp}, \citenamefont {Lalescu},\ and\ \citenamefont
  {Wilczek}}]{Bentkamp2019}%
  \BibitemOpen
  \bibfield  {author} {\bibinfo {author} {\bibfnamefont {L.}~\bibnamefont
  {Bentkamp}}, \bibinfo {author} {\bibfnamefont {C.~C.}\ \bibnamefont
  {Lalescu}},\ and\ \bibinfo {author} {\bibfnamefont {M.}~\bibnamefont
  {Wilczek}},\ }\bibfield  {title} {\bibinfo {title} {Persistent accelerations
  disentangle {L}agrangian turbulence},\ }\href
  {https://doi.org/10.1038/s41467-019-11060-9} {\bibfield  {journal} {\bibinfo
  {journal} {Nat. Commun.}\ }\textbf {\bibinfo {volume} {10}},\ \bibinfo
  {pages} {3550} (\bibinfo {year} {2019})}\BibitemShut {NoStop}%
\bibitem [{\citenamefont {Yamani}\ \emph {et~al.}(2021)\citenamefont {Yamani},
  \citenamefont {Keshavarz}, \citenamefont {Raj}, \citenamefont {Zaki},
  \citenamefont {McKinley},\ and\ \citenamefont
  {Bischofberger}}]{PhysRevLett.127.074501}%
  \BibitemOpen
  \bibfield  {author} {\bibinfo {author} {\bibfnamefont {S.}~\bibnamefont
  {Yamani}}, \bibinfo {author} {\bibfnamefont {B.}~\bibnamefont {Keshavarz}},
  \bibinfo {author} {\bibfnamefont {Y.}~\bibnamefont {Raj}}, \bibinfo {author}
  {\bibfnamefont {T.~A.}\ \bibnamefont {Zaki}}, \bibinfo {author}
  {\bibfnamefont {G.~H.}\ \bibnamefont {McKinley}},\ and\ \bibinfo {author}
  {\bibfnamefont {I.}~\bibnamefont {Bischofberger}},\ }\bibfield  {title}
  {\bibinfo {title} {Spectral universality of elastoinertial turbulence},\
  }\href {https://doi.org/10.1103/PhysRevLett.127.074501} {\bibfield  {journal}
  {\bibinfo  {journal} {Phys. Rev. Lett.}\ }\textbf {\bibinfo {volume} {127}},\
  \bibinfo {pages} {074501} (\bibinfo {year} {2021})}\BibitemShut {NoStop}%
\bibitem [{\citenamefont {Vallis}(2017)}]{vallis2017atmospheric}%
  \BibitemOpen
  \bibfield  {author} {\bibinfo {author} {\bibfnamefont {G.~K.}\ \bibnamefont
  {Vallis}},\ }\href@noop {} {\emph {\bibinfo {title} {Atmospheric and oceanic
  fluid dynamics}}}\ (\bibinfo  {publisher} {Cambridge University Press},\
  \bibinfo {year} {2017})\BibitemShut {NoStop}%
\bibitem [{\citenamefont {Wang}\ \emph {et~al.}(2014)\citenamefont {Wang},
  \citenamefont {Yang},\ and\ \citenamefont {Zhao}}]{C3LC51403J}%
  \BibitemOpen
  \bibfield  {author} {\bibinfo {author} {\bibfnamefont {G.~R.}\ \bibnamefont
  {Wang}}, \bibinfo {author} {\bibfnamefont {F.}~\bibnamefont {Yang}},\ and\
  \bibinfo {author} {\bibfnamefont {W.}~\bibnamefont {Zhao}},\ }\bibfield
  {title} {\bibinfo {title} {There can be turbulence in microfluidics at low
  {R}eynolds number},\ }\href {https://doi.org/10.1039/C3LC51403J} {\bibfield
  {journal} {\bibinfo  {journal} {Lab Chip}\ }\textbf {\bibinfo {volume}
  {14}},\ \bibinfo {pages} {1452} (\bibinfo {year} {2014})}\BibitemShut
  {NoStop}%
\bibitem [{\citenamefont {Di~Carlo}(2009)}]{B912547G}%
  \BibitemOpen
  \bibfield  {author} {\bibinfo {author} {\bibfnamefont {D.}~\bibnamefont
  {Di~Carlo}},\ }\bibfield  {title} {\bibinfo {title} {Inertial
  microfluidics},\ }\href {https://doi.org/10.1039/B912547G} {\bibfield
  {journal} {\bibinfo  {journal} {Lab Chip}\ }\textbf {\bibinfo {volume} {9}},\
  \bibinfo {pages} {3038} (\bibinfo {year} {2009})}\BibitemShut {NoStop}%
\bibitem [{\citenamefont {Zhang}\ \emph {et~al.}(2016)\citenamefont {Zhang},
  \citenamefont {Yan}, \citenamefont {Yuan}, \citenamefont {Alici},
  \citenamefont {Nguyen}, \citenamefont {Ebrahimi~Warkiani},\ and\
  \citenamefont {Li}}]{C5LC01159K}%
  \BibitemOpen
  \bibfield  {author} {\bibinfo {author} {\bibfnamefont {J.}~\bibnamefont
  {Zhang}}, \bibinfo {author} {\bibfnamefont {S.}~\bibnamefont {Yan}}, \bibinfo
  {author} {\bibfnamefont {D.}~\bibnamefont {Yuan}}, \bibinfo {author}
  {\bibfnamefont {G.}~\bibnamefont {Alici}}, \bibinfo {author} {\bibfnamefont
  {N.-T.}\ \bibnamefont {Nguyen}}, \bibinfo {author} {\bibfnamefont
  {M.}~\bibnamefont {Ebrahimi~Warkiani}},\ and\ \bibinfo {author}
  {\bibfnamefont {W.}~\bibnamefont {Li}},\ }\bibfield  {title} {\bibinfo
  {title} {Fundamentals and applications of inertial microfluidics: a review},\
  }\href {https://doi.org/10.1039/C5LC01159K} {\bibfield  {journal} {\bibinfo
  {journal} {Lab Chip}\ }\textbf {\bibinfo {volume} {16}},\ \bibinfo {pages}
  {10} (\bibinfo {year} {2016})}\BibitemShut {NoStop}%
\bibitem [{\citenamefont {Roy}\ \emph {et~al.}(2003)\citenamefont {Roy},
  \citenamefont {Raju}, \citenamefont {Chuang}, \citenamefont {Cruden},\ and\
  \citenamefont {Meyyappan}}]{doi:10.1063/1.1559936}%
  \BibitemOpen
  \bibfield  {author} {\bibinfo {author} {\bibfnamefont {S.}~\bibnamefont
  {Roy}}, \bibinfo {author} {\bibfnamefont {R.}~\bibnamefont {Raju}}, \bibinfo
  {author} {\bibfnamefont {H.~F.}\ \bibnamefont {Chuang}}, \bibinfo {author}
  {\bibfnamefont {B.~A.}\ \bibnamefont {Cruden}},\ and\ \bibinfo {author}
  {\bibfnamefont {M.}~\bibnamefont {Meyyappan}},\ }\bibfield  {title} {\bibinfo
  {title} {Modeling gas flow through microchannels and nanopores},\ }\href
  {https://doi.org/10.1063/1.1559936} {\bibfield  {journal} {\bibinfo
  {journal} {J. Appl. Phys.}\ }\textbf {\bibinfo {volume} {93}},\ \bibinfo
  {pages} {4870} (\bibinfo {year} {2003})}\BibitemShut {NoStop}%
\bibitem [{\citenamefont {Straube}\ \emph {et~al.}(2020)\citenamefont
  {Straube}, \citenamefont {Kowalik}, \citenamefont {Netz},\ and\ \citenamefont
  {H{\"o}fling}}]{Straube2020}%
  \BibitemOpen
  \bibfield  {author} {\bibinfo {author} {\bibfnamefont {A.~V.}\ \bibnamefont
  {Straube}}, \bibinfo {author} {\bibfnamefont {B.~G.}\ \bibnamefont
  {Kowalik}}, \bibinfo {author} {\bibfnamefont {R.~R.}\ \bibnamefont {Netz}},\
  and\ \bibinfo {author} {\bibfnamefont {F.}~\bibnamefont {H{\"o}fling}},\
  }\bibfield  {title} {\bibinfo {title} {Rapid onset of molecular friction in
  liquids bridging between the atomistic and hydrodynamic pictures},\ }\href
  {https://doi.org/10.1038/s42005-020-0389-0} {\bibfield  {journal} {\bibinfo
  {journal} {Commun. Phys.}\ }\textbf {\bibinfo {volume} {3}},\ \bibinfo
  {pages} {126} (\bibinfo {year} {2020})}\BibitemShut {NoStop}%
\bibitem [{\citenamefont {Kavokine}\ \emph {et~al.}(2021)\citenamefont
  {Kavokine}, \citenamefont {Netz},\ and\ \citenamefont
  {Bocquet}}]{doi:10.1146/annurev-fluid-071320-095958}%
  \BibitemOpen
  \bibfield  {author} {\bibinfo {author} {\bibfnamefont {N.}~\bibnamefont
  {Kavokine}}, \bibinfo {author} {\bibfnamefont {R.~R.}\ \bibnamefont {Netz}},\
  and\ \bibinfo {author} {\bibfnamefont {L.}~\bibnamefont {Bocquet}},\
  }\bibfield  {title} {\bibinfo {title} {Fluids at the nanoscale: From
  continuum to subcontinuum transport},\ }\href
  {https://doi.org/10.1146/annurev-fluid-071320-095958} {\bibfield  {journal}
  {\bibinfo  {journal} {Annu. Rev. Fluid Mech.}\ }\textbf {\bibinfo {volume}
  {53}},\ \bibinfo {pages} {377} (\bibinfo {year} {2021})}\BibitemShut
  {NoStop}%
\bibitem [{\citenamefont {Coussot}(2017)}]{coussot2017mudflow}%
  \BibitemOpen
  \bibfield  {author} {\bibinfo {author} {\bibfnamefont {P.}~\bibnamefont
  {Coussot}},\ }\href@noop {} {\emph {\bibinfo {title} {Mudflow rheology and
  dynamics}}}\ (\bibinfo  {publisher} {Routledge},\ \bibinfo {year}
  {2017})\BibitemShut {NoStop}%
\bibitem [{\citenamefont {Brenner}\ \emph {et~al.}(2002)\citenamefont
  {Brenner}, \citenamefont {Hilgenfeldt},\ and\ \citenamefont
  {Lohse}}]{RevModPhys.74.425}%
  \BibitemOpen
  \bibfield  {author} {\bibinfo {author} {\bibfnamefont {M.~P.}\ \bibnamefont
  {Brenner}}, \bibinfo {author} {\bibfnamefont {S.}~\bibnamefont
  {Hilgenfeldt}},\ and\ \bibinfo {author} {\bibfnamefont {D.}~\bibnamefont
  {Lohse}},\ }\bibfield  {title} {\bibinfo {title} {Single-bubble
  sonoluminescence},\ }\href {https://doi.org/10.1103/RevModPhys.74.425}
  {\bibfield  {journal} {\bibinfo  {journal} {Rev. Mod. Phys.}\ }\textbf
  {\bibinfo {volume} {74}},\ \bibinfo {pages} {425} (\bibinfo {year}
  {2002})}\BibitemShut {NoStop}%
\bibitem [{\citenamefont {Toegel}\ \emph {et~al.}(2006)\citenamefont {Toegel},
  \citenamefont {Luther},\ and\ \citenamefont {Lohse}}]{PhysRevLett.96.114301}%
  \BibitemOpen
  \bibfield  {author} {\bibinfo {author} {\bibfnamefont {R.}~\bibnamefont
  {Toegel}}, \bibinfo {author} {\bibfnamefont {S.}~\bibnamefont {Luther}},\
  and\ \bibinfo {author} {\bibfnamefont {D.}~\bibnamefont {Lohse}},\ }\bibfield
   {title} {\bibinfo {title} {Viscosity destabilizes sonoluminescing bubbles},\
  }\href {https://doi.org/10.1103/PhysRevLett.96.114301} {\bibfield  {journal}
  {\bibinfo  {journal} {Phys. Rev. Lett.}\ }\textbf {\bibinfo {volume} {96}},\
  \bibinfo {pages} {114301} (\bibinfo {year} {2006})}\BibitemShut {NoStop}%
\bibitem [{\citenamefont {Zakari}\ and\ \citenamefont
  {Jou}(1993)}]{PhysRevD.48.1597}%
  \BibitemOpen
  \bibfield  {author} {\bibinfo {author} {\bibfnamefont {M.}~\bibnamefont
  {Zakari}}\ and\ \bibinfo {author} {\bibfnamefont {D.}~\bibnamefont {Jou}},\
  }\bibfield  {title} {\bibinfo {title} {Equations of state and transport
  equations in viscous cosmological models},\ }\href
  {https://doi.org/10.1103/PhysRevD.48.1597} {\bibfield  {journal} {\bibinfo
  {journal} {Phys. Rev. D}\ }\textbf {\bibinfo {volume} {48}},\ \bibinfo
  {pages} {1597} (\bibinfo {year} {1993})}\BibitemShut {NoStop}%
\bibitem [{\citenamefont {Jung}\ \emph {et~al.}(2017)\citenamefont {Jung},
  \citenamefont {Hanke},\ and\ \citenamefont {Schmid}}]{jung2017iterative}%
  \BibitemOpen
  \bibfield  {author} {\bibinfo {author} {\bibfnamefont {G.}~\bibnamefont
  {Jung}}, \bibinfo {author} {\bibfnamefont {M.}~\bibnamefont {Hanke}},\ and\
  \bibinfo {author} {\bibfnamefont {F.}~\bibnamefont {Schmid}},\ }\bibfield
  {title} {\bibinfo {title} {Iterative reconstruction of memory kernels},\
  }\href {https://doi.org/10.1021/acs.jctc.7b00274} {\bibfield  {journal}
  {\bibinfo  {journal} {J. Chem. Theory Comput.}\ }\textbf {\bibinfo {volume}
  {13}},\ \bibinfo {pages} {2481} (\bibinfo {year} {2017})}\BibitemShut
  {NoStop}%
\bibitem [{\citenamefont {Meyer}\ \emph {et~al.}(2020)\citenamefont {Meyer},
  \citenamefont {Pelagejcev},\ and\ \citenamefont {Schilling}}]{meyer2020non}%
  \BibitemOpen
  \bibfield  {author} {\bibinfo {author} {\bibfnamefont {H.}~\bibnamefont
  {Meyer}}, \bibinfo {author} {\bibfnamefont {P.}~\bibnamefont {Pelagejcev}},\
  and\ \bibinfo {author} {\bibfnamefont {T.}~\bibnamefont {Schilling}},\
  }\bibfield  {title} {\bibinfo {title} {Non-markovian out-of-equilibrium
  dynamics: A general numerical procedure to construct time-dependent memory
  kernels for coarse-grained observables},\ }\href
  {https://doi.org/10.1209/0295-5075/128/40001} {\bibfield  {journal} {\bibinfo
   {journal} {Europhys. Lett.}\ }\textbf {\bibinfo {volume} {128}},\ \bibinfo
  {pages} {40001} (\bibinfo {year} {2020})}\BibitemShut {NoStop}%
\bibitem [{\citenamefont {Lesnicki}\ \emph {et~al.}(2016)\citenamefont
  {Lesnicki}, \citenamefont {Vuilleumier}, \citenamefont {Carof},\ and\
  \citenamefont {Rotenberg}}]{PhysRevLett.116.147804}%
  \BibitemOpen
  \bibfield  {author} {\bibinfo {author} {\bibfnamefont {D.}~\bibnamefont
  {Lesnicki}}, \bibinfo {author} {\bibfnamefont {R.}~\bibnamefont
  {Vuilleumier}}, \bibinfo {author} {\bibfnamefont {A.}~\bibnamefont {Carof}},\
  and\ \bibinfo {author} {\bibfnamefont {B.}~\bibnamefont {Rotenberg}},\
  }\bibfield  {title} {\bibinfo {title} {Molecular hydrodynamics from memory
  kernels},\ }\href {https://doi.org/10.1103/PhysRevLett.116.147804} {\bibfield
   {journal} {\bibinfo  {journal} {Phys. Rev. Lett.}\ }\textbf {\bibinfo
  {volume} {116}},\ \bibinfo {pages} {147804} (\bibinfo {year}
  {2016})}\BibitemShut {NoStop}%
\bibitem [{\citenamefont {Daldrop}\ \emph {et~al.}(2017)\citenamefont
  {Daldrop}, \citenamefont {Kowalik},\ and\ \citenamefont
  {Netz}}]{PhysRevX.7.041065}%
  \BibitemOpen
  \bibfield  {author} {\bibinfo {author} {\bibfnamefont {J.~O.}\ \bibnamefont
  {Daldrop}}, \bibinfo {author} {\bibfnamefont {B.~G.}\ \bibnamefont
  {Kowalik}},\ and\ \bibinfo {author} {\bibfnamefont {R.~R.}\ \bibnamefont
  {Netz}},\ }\bibfield  {title} {\bibinfo {title} {External potential modifies
  friction of molecular solutes in water},\ }\href
  {https://doi.org/10.1103/PhysRevX.7.041065} {\bibfield  {journal} {\bibinfo
  {journal} {Phys. Rev. X}\ }\textbf {\bibinfo {volume} {7}},\ \bibinfo {pages}
  {041065} (\bibinfo {year} {2017})}\BibitemShut {NoStop}%
\end{thebibliography}
\end{document}


\title{
       Supplementary Material\\Shear and bulk acceleration viscosities in simple fluids}

\author{Johannes Renner}
\affiliation{Theoretische Physik II, Physikalisches Institut,
  Universit{\"a}t Bayreuth, D-95440 Bayreuth, Germany}
\author{Matthias Schmidt}
\affiliation{Theoretische Physik II, Physikalisches Institut,
  Universit{\"a}t Bayreuth, D-95440 Bayreuth, Germany}
 \author{Daniel de las Heras}
\affiliation{Theoretische Physik II, Physikalisches Institut,
  Universit{\"a}t Bayreuth, D-95440 Bayreuth, Germany}

\date{\today}

\maketitle
The Supplementary Material contains the expression for the temporal contribution of the current,
(\!\ref{1}), the description of the calculation of the kernel parameters (\!\ref{2}),
details about molecular (\!\ref{3}) and Brownian (\!\ref{4}) dynamics simulations,
summaries of the custom flow method (\!\ref{5}), the splitting of internal forces
into viscous and structural contributions (\!\ref{6}) and power functional theory (\!\ref{7}),
as well as supplementary data on the shear flow in Brownian dynamics (\!\ref{8}) and on
the traveling shear wave (\!\ref{9}).

\subsection{Time evolution of the current}\label{1}
The temporal contribution to the current, shown in Fig.1(a) of the main text, is set to 
\begin{align}
J_t(t)=
    \begin{cases}
            0.5\left[1 - \cos\left({\pi t}/{t_\uparrow}\right)\right],& \qquad 0<t\leq t_\uparrow \\
            1        ,& \qquad t_\uparrow<t\leq t_c\\
            0.5\left[1+\cos\left(\pi\frac{t-t_c}{t_\downarrow-t_c}\right)\right],& \qquad t_c<t\leq t_\downarrow\\
            0        ,& \qquad t_\downarrow<t, 
    \end{cases}\label{eq:timeramp}
\end{align}
with  $t_\uparrow=1\tau$, $t_c=5\tau$, and $t_\downarrow=6\tau$.
\subsection{Calculation of the kernel parameters}\label{2}
To obtain the memory times and the amplitudes of the viscosity kernels we proceed as follows.
For the bulk and the shear flows considered here, the one-body density is by construction time-independent $\rho(\rv,t)=\rho(\rv)$ and the one-body current $\J$ factorizes into time- and space-dependent parts $\J(\rv,t)=\J_{\rv}(\rv)J_t(t)$.
Hence, the time derivative of the current also factorizes into time- and space-dependent parts 
\begin{equation}
\Jdot(\rv,t)=\J_{\rv}(\rv)\dot{J}_t(t),
\end{equation}
and it has the same spatial form $\J_{\rv}(\rv)$ as the current itself.
Since the expressions for the shear $\fs$ and bulk $\fb$ viscous forces are linear in both $\vel=\J/\rho=J_t\J_{\rv}/\rho$ and $\ac=\dot\vel$ where here $\dot\vel=\Jdot/\rho=\dot{J}_t\J_{\rv}/\rho$ [see Eqs. (1) and (2) of the main text],
the viscous forces also factorize into time- and space-dependent parts:
\begin{align}
	\textbf{f}_\alpha(\rv,t)=C_\alpha(t) \textbf{f}_{\rv,\alpha}(\rv),\;\;\;\;\alpha=b,s\label{eq:viscoustimespacesplit},
\end{align}
where the space-dependent parts are
\begin{eqnarray}
        \textbf{f}_{\rv,b} & = & \frac{1}{\rho}\nabla\left[\rho\rho\nabla\cdot\left(\dfrac{\J_{\rv}}{\rho}\right)\right],\label{eqfvissb}\\
        \textbf{f}_{\rv,s} & = &-\frac{1}{\rho}\nabla\times\left[\rho\rho\nabla\times\left(\dfrac{\J_{\rv}}{\rho}\right)\right],\label{eqfvisss}
\end{eqnarray}
and the temporal parts are
\begin{equation}
	C_\alpha(t)=\int\limits_0^t dt'\left(K_\alpha^{\vel}(t-t')J_t(t')+K_\alpha^{\acc}(t-t')\dot{J}_t(t')\right),\label{calpha}
\end{equation}
with $\alpha=b$ for bulk and $\alpha=s$ for shear. The kernels are
\begin{equation}
	K_\alpha^{\boldsymbol\Gamma}(t)=\frac{c_\alpha^{\boldsymbol\Gamma}}{\tau_\alpha^{\boldsymbol\Gamma}}\exp(-t/\tau_\alpha^{\boldsymbol\Gamma}),
\end{equation}
with the superscript $\boldsymbol\Gamma$ labeling either the acceleration $\boldsymbol\Gamma=\acc$ or the velocity $\boldsymbol\Gamma=\vel$ contributions.
The factorization of the viscous force into temporal- and spatial parts, Eq.~\eqref{eq:viscoustimespacesplit}, which facilitates the analysis of the data, is not general and holds only if the one-body current also factorizes. 
Custom flow is therefore an essential tool here since it allows to carefully prescribe the features of the flow.
We show in Supplementary Fig.~\ref{fig:Sup1} the space-dependent parts of the bulk and shear viscous forces at different times according to simulations. As expected, the curves for different times collapse into a single curve.

The process to calculate the kernel parameters uses two steps. In step one,
at every time $t$ we compare the simulation data for $\fvis$ to Eq.~\eqref{eq:viscoustimespacesplit} using the expressions in Eqs.~\eqref{eqfvissb} and~\eqref{eqfvisss} for the spatial part of the viscous forces.
As a result, we obtain the curve $C_\alpha(t)$ in simulations.
In the second step, the kernel parameters are obtained by finding the values of $c_\alpha^{\boldsymbol\Gamma}$ and $\tau_\alpha^{\boldsymbol\Gamma}$ in Eq.~\eqref{calpha} that best reproduce the curve $C_\alpha(t)$ that results from step one.

\begin{figure}
    \centering
    \resizebox{0.5\textwidth}{!}{\includegraphics{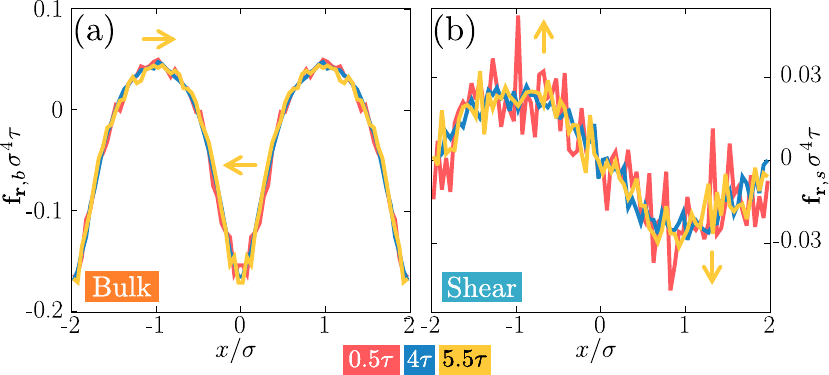}}
        \caption{
		Space-dependent parts of (a) the bulk $\textbf{f}_{\rv,b}$ and (b) the shear $\textbf{f}_{\rv,s}$ viscous forces as a function of the $x$-coordinate obtained with MD simulation data. Three different times are shown, as indicated by the color of the lines. The arrows indicate the direction of the vector field at the selected positions.
            }
    \label{fig:Sup1}
\end{figure}

\subsection{Molecular dynamics simulations}\label{3}
We use molecular dynamics (MD) to simulate a dynamical ensemble of $\sim10^6$ instances of a three-dimensional system consisting of $N=50$ particles interacting via the purely repulsive Weeks-Chandler-Andersen interparticle-interaction potential~\cite{WCA-potential}
\begin{align}
    \phi(r_{ij})= 
\begin{cases}
    4\epsilon\left[\left(\frac{\sigma}{r_{ij}}\right)^{12}-\left(\frac{\sigma}{r_{ij}}\right)^6\right]& \text{if } r_{ij}\geq r_c\\
    0&              \text{otherwise}.
\end{cases}
\end{align}
Here, $r_{ij}=|\textbf{r}_i-\textbf{r}_j|$ is the distance between particle $i$ and $j$, and $r_c=2^\frac{1}{6}\,\sigma$ is the cutoff radius, which is located at the minimum of the Lennard-Jones potential.
We work in units of the length scale $\sigma$, the energy scale $\epsilon$, and the mass of one particle $m$. Hence, the derived time scale is $\tau=\sqrt{m\sigma^2/\epsilon}$.

The equations of motion for the $i$th particle are
\begin{eqnarray}
        \frac{d\textbf{r}_i}{dt}&=&\frac{\textbf{p}_i}{m},\label{eq:N1}\\
        \frac{d\textbf{p}_i}{dt}&=&-\nabla_iu(\textbf{r}^N)+\fext(\textbf{r}_i,t),\label{eq:N2}
\end{eqnarray}
where $\textbf{r}_i$ denotes the position of the $i$th particle, and $\textbf{p}_i=m\textbf{v}_i$ its momentum, with $\textbf{v}_i$ its
velocity.
The total force acting on the particle is made of an external contribution $\fext(\textbf{r}_i,t)$, and an internal one, $-\nabla_iu(\textbf{r}^N)$.
Here, $\nabla_i$ is the partial derivative with respect to $\textbf{r}_i$ and
$u(\textbf{r}^N)=\frac12\sum_i\sum_{j\neq i}\phi(r_{ij})$ is the total interparticle potential energy with, $\textbf{r}^N=\{\textbf{r}_1\dots\textbf{r}_N\}$
the complete set of particle positions.

We integrate the many-body equations of motion in MD using the standard velocity-Verlet algorithm with time step $dt=10^{-4}\tau$.
The simulation box is a cuboid with lengths $L_x=4\sigma$, $L_y=10\sigma$ and $L_z=8\sigma$ and periodic boundary conditions.
To spatially resolve the one-body fields we discretize the system in the $x$-coordinate with bins of size ${0.05}{\sigma}$.

The particle positions are initialised randomly with the constraint that no interparticle interaction occur.
The particle velocities are initialised following a Maxwell-Boltzmann distribution with absolute temperature $T$.
For the initial equilibration of the shear flow (homogeneous density) we let the system evolve for $1\,\tau$ without external force.
To initialize the compressible flow (inhomogeneous density profile), we use custom flow to grow the density inhomogeneity and then let the system equilibrate for $4\,\tau$ such that memory effects decay.
The starting temperature, calculated from the kinetic energy using the equipartition theorem, is set to $k_BT/\epsilon=0.59$ (compressible flow) and $k_BT/\epsilon=0.486$ (shear flow).
Here,  $k_B$ is the Boltzmann constant. The temperatures of the final equilibrium states are $k_BT/\epsilon=0.60$ and $k_BT/\epsilon=0.492$ for the compressible and the shear flows, respectively. These values are slightly higher than the initial values due to the heating induced by the external driving. Since the temperature increase was small (below $2\%$) we did not use a thermostat. Note however that custom flow can also be implemented together with a thermostat~\cite{customflowMD}.

The one-body fields of interest are resolved in space and in time. For example, the one-body density and current profiles are given by
\begin{eqnarray}
	\rho(\textbf{r},t)&=&\left\langle\sum_{i=1}^N\delta\left(\textbf{r}-\textbf{r}_i\right)\right\rangle,\label{eq:density}\\
	\textbf{J}(\textbf{r},t)&=&\left\langle\sum_{i=1}^N\delta(\textbf{r}-\textbf{r}_i)\textbf{v}_i\right\rangle,\label{eq:current}
\end{eqnarray}
with $\delta(\textbf{r})$ being the three dimensional Dirac delta distribution, and $\textbf{r}$ being the position vector.
The statistical average, denoted by the brackets $\langle\cdot\rangle$ 
is done at each time $t$ over different realizations of the initial conditions (the positions and the velocities of the particles at
the initial time $t=0$). Specifically, we average over $2\cdot 10^6$ different realizations (initial states).

\subsection{Brownian dynamics simulations}\label{4}
For the overdamped Brownian dynamics simulations we use the standard Euler algorithm to integrate the equation of motion of the $i$th particle
\begin{align}
\rv_i(t+dt)=\rv_i(t)+\frac{dt}{\gamma}
   [-\nabla_i u(\rv^N)
  + {\bf f}_{\rm ext}(\rv_i,t)]
 + \boldsymbol\eta_i(t),
\label{EQBD}
\end{align}
where $\boldsymbol\eta_i$ is a delta-correlated Gaussian random
displacement with standard deviation $\sqrt{2dt k_BT/\gamma}$ in
accordance with the fluctuation-dissipation theorem and $\gamma$ is the friction coefficient against the (implicit) solvent.
We hence use in Eq.~\eqref{EQBD} the standard assumption that the random force does not depend on the external force~\cite{Kubo_1966}.
The integration time step is set to $dt = 10^{-4}\tau_b$
with $\tau_b=\sigma^2\gamma/\epsilon$ the BD unit of time.
In BD we work in units of $\sigma$, $\epsilon$, and $\gamma$.
We average over $4\cdot 10^6$ trajectories, i.e. twice than in MD, due to the larger statistical noise generated by the random force.
The velocity of particle $i$ at time $t$, required to e.g. sample the current following Eq.~\eqref{eq:current}, is calculated as the central derivative of the position vector~\cite{de2019custom}:
\begin{equation}
	\vel_i(t)=\frac{\rv_i(t+dt)-\rv_i(t-dt)}{2dt}.
\end{equation}

All further parameters of the simulation, i.e. temperature, number of particles, and target fields, are the same as in MD.

Since the external driving is time-dependent, the overdamped approximation that underlies Eq.~\eqref{EQBD} might not be accurate.
However, we use here overdamped BD only as a reference system in which inertial effects are eliminated by construction. 
This allows us to highlight the inertial effects that occur in MD.

\subsection{Custom flow}\label{5}

Custom flow is a numerical method that finds the external force corresponding to prescribed density, velocity, and acceleration fields (the target fields).
A complete description of the method is given in Refs.~\cite{customflowMD,de2019custom}. Here, we only summarize the main ideas of custom flow in molecular dynamics.
The external force is found iteratively. At each iteration, the external force is the same as in the previous iteration plus a term that aims to correct the differences between sampled and target fields,
\begin{equation}
	\fext^{(k+1)}(\rv,t)=\fext^{(k)}(\rv,t)+\dfrac{m}{\rho(\rv,t)\Delta t}\left(\textbf{J}(\rv,t)-\textbf{J}^{(k)}(\rv,t)\right).\label{eq:MDcustomflow}
\end{equation}
Here, $k$ is the iteration index. Hence, $\fext^{(k)}(\rv,t)$ and $\textbf{J}^{(k)}(\rv,t)$ denote the external force and the current sampled at iteration $k$, whereas $\rho(\rv,t)$ and $\J(\rv,t)$ are
the target fields. The convergence of the iteration scheme~\eqref{eq:MDcustomflow} is achieved when the external forces at iterations $k+1$ and $k$ coincide within a given tolerance (in practice less than ten iterations are usually enough to achieve convergence).
The whole iteration scheme needs to be repeated at time intervals separated by $\Delta t$ which we set to be $\Delta t = 10 dt$, i.e., ten times bigger than the time step of the simulation $dt$. 
At each time, we initialize the external force according to
\begin{equation}
	\fext^{(0)}(\rv,t)=\frac{m\Jdot(\rv,t)}{\rho(\rv,t)},
\end{equation}
which follows from the exact one-body force balance equation~\eqref{eq:MDonebodyforcebalance} by making the internal force $\fint$ and the transport term $\divtau$ zero everywhere. 

Using Eq.~(\ref{eq:MDcustomflow}), custom flow MD minimizes the difference between target and sampled one-body currents. This results in very accurate (essentially noise free) sampled currents. The noise, which in standard MD simulations usually occurs in the sampled fields, appears in custom flow in the external force which is tailored to the initial set of microstates (we use $2\cdot 10^6$ different initial states), see Supplementary Figure~\ref{fig:MDrawexternal}. For a better visual representation we show in the main paper and also in Supplementary Figure~\ref{fig:MDrawexternal} smooth external force profiles which result from removing the high Fourier modes of the raw signal. Both, the external force that follows directly from custom flow and its smoothed version produce very similar dynamics~\cite{customflowMD}.

\begin{figure}
    \centering
    \resizebox{0.5\textwidth}{!}{\includegraphics{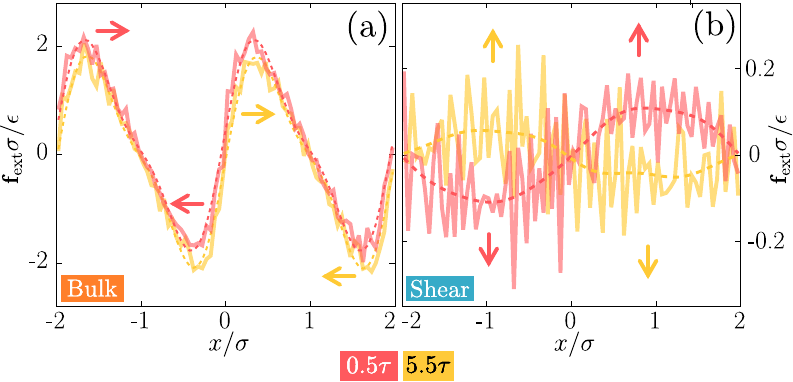}}
	\caption{External force $\fext$ produced by MD custom flow (solid thick lines) as a function of $x$ for (a) bulk and (b) shear flows at times $0.5\tau$ (red) and $5.5\tau$ (yellow).
	The smoothed external forces, obtained by removing the high frequency modes, are also shown with dashed lines.
	Custom flow minimizes the statistical noise that usually occurs in the sampled fields like the density and the velocity profiles.
	As a result, the external forces obtained with custom flow appear to be noisy.
	The colored arrows indicate the direction of the force at the selected positions.}
    \label{fig:MDrawexternal}
\end{figure}

\subsection{Viscous and structural internal forces}\label{sec:forcesplitting}\label{6}
In non-equilibrium, the total internal force $\fint$, which is solely generated by the interparticle interactions, contains structural and flow contributions~\cite{de2020flow}.
The structural part is able to e.g. sustain gradients in the density profile, whereas the flow contribution represents the viscous response of the system.
The total internal force is easily accessible in computer simulations since $\fint(\rv,t)=\Fint(\rv,t)/\rho(\rv,t)$ with $\Fint$ being the internal force density
\begin{equation}
\Fint(\textbf{r},t)=-\left\langle\sum_{i=1}^N\delta(\textbf{r}-\textbf{r}_i)
        \nabla_i u\left(\textbf{r}^N\right)
        \right\rangle.\label{eq:Fint}
\end{equation}
To extract the viscous forces from the total internal force, we use that the viscous forces are sensitive to the direction of the flow.
Hence, reversing the direction of the flow, i.e.,  $\vel\rightarrow-\vel$ and $\acc\rightarrow-\acc$ while keeping the density profile unchanged, flips the sign of the viscous forces and leaves the structural forces unchanged~\cite{de2020flow}.
The sign change of $\fvis$ by reversing the direction of the flow is apparent in Eqs. (1) and (2) of the main text.

Hence, the total viscous force of the system can be calculated as~\cite{de2020flow}
\begin{equation}
	\fvis(\rv,t)=\dfrac{\fint(\rv,t)-\fint^{\,\,r}(\rv,t)}{2},\label{eq:reversal}
\end{equation}
where $\fint^{\,\,r}(\rv,t)$ indicates the internal force in the reverse system, i.e. a system with flow velocity $-\vel(\rv,t)$, acceleration $-\acc(\rv,t)$, but the same density profile $\rho(\rv,t)$ as the original forward system [in which the flow
is given by $+\vel(\rv,t)$ and $+\acc(\rv,t)$ and the internal force is $\fint(\rv,t)$].

Using Eq.~\eqref{eq:reversal} to measure the viscous force is always possible if the density profile is time-independent, such as e.g. steady states and the full non-equilibrium flows designed here. 
If the density varies in time, then temporal changes of the density profile affect the flow (via the continuity equation).
In such cases finding the reverse system to unambiguously measure the viscous force is in general not possible.
This again highlights the importance of custom flow that allows us to generate flows in which the viscous response can be unambiguously measured.

To create the reverse system we follow two independent methods which give the same results. The first method simply uses custom flow to prescribe the respective reverse flow and find the corresponding external force.
The second method makes use of symmetry arguments to compute the reverse flow from the forward flow and hence obtain $\fint^{\,\,r}$ from $\fint$.
This second possibility, which we describe in detail in what follows, is possible only due to the specific characteristics of the flows.
In the general case the first method is required to find the reverse system. 

{\bf Bulk flow.} Let us consider a virtual flow in which we reverse at each time $t$ only the $x$-component of every particle in the original flow, i.e.\ we perform the operation $x_i(t)\rightarrow-x_i(t)$ while keeping the other two components unchanged.
Hence, the $x$-component of the current in this virtual system, $J_x^v$, is
\begin{eqnarray}
	J_x^v&=&\left\langle\sum\limits_{i=1}^N\delta(x-(-x_i))\dfrac{d(-x_i)}{dt}\right\rangle\nonumber\\
	&=&-\left\langle\sum\limits_{i=1}^N\delta(x+x_i)v^x_i\right\rangle=-J_x(-x,t).\label{bulkJx}
\end{eqnarray}
By construction, the bulk flow has the symmetry $J_x(x,t)=J_x(-x,t)$. Hence, in combination with Eq.~\eqref{bulkJx} above we conclude that $J_x^v(x,t)=J_x^r(x,t)$.
Also by construction, the other components of the current vanish and the density profile has also the same symmetry $\rho(x,t)=\rho(-x,t)$.
Therefore, we can construct the reverse system of the bulk flow by simply using the trajectories of the forward system and performing the operation $x_i(t)\rightarrow-x_i(t)$.

Hence, for the bulk flow the $x$-component of the internal force density in the reverse system is
\begin{align}
F_\text{int,x}^r(x,t)&=\left\langle\sum\limits_{i=1}^N\delta(x+x_i)\dfrac{\partial\phi(r_{ij})}{\partial (-x_i)}\right\rangle\\
&=-\left\langle\sum\limits_{i=1}^N\delta(x+x_i)\dfrac{\partial\phi(r_{ij})}{\partial x_i}\right\rangle\\
&=-F_\text{int,x}(-x,t),
\end{align}
where we have used that the interparticle distance $r_{ij}$ is not affected by the transformation $x_i\rightarrow-x_i$.
Due to the spatial symmetry of the density profile $\rho(x,t)=\rho(-x,t)$, the internal force $\fint$ has the same symmetry as the internal force density $\Fint$, i.e. $f_\text{int,x}^r(x,t)=-f_\text{int,x}^r(-x,t)$ because $\Fint=\fint/\rho$.
Therefore, for the bulk flow the viscous part of the total internal force, see Eq.~\eqref{eq:reversal}, can be obtained from the forward bulk flow as a simple arithmetic mean
\begin{align}
	f_{\text{vis},x}(x,t)=\dfrac{f_{\text{int},x}(x,t)+f_{\text{int},x}(-x,t)}{2}.
\end{align}

{\bf Shear flow.} Here, the flow is directed along the $y$-axis and the density is homogeneous $\nabla\rho=0$.
Therefore, by construction, the $y$-component of the internal force is only of viscous nature (no structural term).
We arrive at the same conclusion by considering a virtual flow in which we reverse at each time the $y$-component of all particles, i.e. $y_i(t)\rightarrow-y_i(t)$.
Hence, the $y$-component of the current in the virtual system $J^v_y$ is 
\begin{align}
    J_y^v(x,t)&=\left\langle\sum\limits_{i=1}^N\delta(x-x_i)\dfrac{d(-y_i)}{dt}\right\rangle\\
    &=-\left\langle\sum\limits_{i=1}^N\delta(x-x_i)v^y_i\right\rangle=-J_y(x,t),
\end{align}
which is precisely the $y$-component of the current in the reverse system $J_y^r(x,t)=J_y^v(x,t)$.
Given that the other two components of the current vanish and that the density profile is stationary, we conclude that the reverse system can be obtained from the forward flow by
simply using the operation $y_i(t)\rightarrow-y_i(t)$ and performing the desired averages.

The $y$-component of the internal force density in the reverse system is therefore 
\begin{align}
F_\text{int,y}^r(x,t)&=\left\langle\sum\limits_{i=1}^N\delta(x-x_i)\dfrac{\partial\phi(r_{ij})}{\partial (-y_i)}\right\rangle\\
&=-\left\langle\sum\limits_{i=1}^N\delta(x-x_i)\dfrac{\partial\phi(r_{ij})}{\partial y_i}\right\rangle\\
&=-F_\text{int,y}(x,t).
\end{align}
Hence, using Eq.~\eqref{eq:reversal}, the viscous part is  
\begin{align}
	f_{\text{vis},y}(x,t)=\dfrac{f_{\text{int},y}(x,t)+f_{\text{int},y}(x,t)}{2}=f_{\text{int},y}(x,t).
\end{align}
That is, as expected, the flow-direction of the internal force in a shear flow contains only viscous terms provided that there is no density inhomogeneity in the flow direction.

\subsection{Power functional theory}\label{7}
Power functional theory (PFT) is a variational theory that describes the dynamics of interacting many-body overdamped~\cite{PowerF} and  inertial~\cite{PFTMD} systems at the level of one-body fields.
A variational principle produces by construction the exact one-body force balance equation of the system. For a classical system of particles following the equations of motion~\eqref{eq:N1} and~\eqref{eq:N2}, the exact one-body force balance equation reads~\cite{PFTMD}
\begin{equation}
    m\Jdot(\textbf{r},t)=\rho(\textbf{r},t)\left[\fext(\textbf{r},t)+\fint(\textbf{r},t)\right]+\divtau(\textbf{r},t),\label{eq:MDonebodyforcebalance}
\end{equation}
where the last term involves the divergence of the second rank kinetic stress tensor $\boldsymbol\tau$ and it describes transport effects that arise due to the one-body description of the dynamics. In thermal equilibrium this term reduces to diffusive transport $\nabla\cdot\boldsymbol\tau=-k_BT\nabla\rho$. In simulations, $\boldsymbol\tau$ can be sampled via
\begin{equation}
	        \boldsymbol{\tau}(\textbf{r},t)=-m\left\langle\sum_{i=1}^N\delta(\textbf{r}-\textbf{r}_i)\textbf{v}_i\textbf{v}_i\right\rangle,\label{eq:tau}
\end{equation}
where $\textbf{v}_i\textbf{v}_i$ indicates the dyadic product of the velocity of particle $i$ with itself.

Within PFT each term of the force balance equation~\eqref{eq:MDonebodyforcebalance} is generated via a functional derivative of a corresponding functional generator with respect to the time derivative of the current or alternatively with respect to the acceleration field.
The density profile $\rho$, the current $\J$ (or the velocity $\vel=\J/\rho$), and the time derivative of the current $\Jdot$ (or the acceleration $\acc=\dot\vel=\Jdot/\rho$, where the second equality holds only if $\dot\rho=0$ like in the present work) are the natural functional dependencies of the generator functionals.
One important task in PFT is to find an approximated functional that generates via functional differentiation the internal force field.

The simplest approximation based on an expansion in terms of the acceleration gradient $\nabla\acc$ that is (i) compatible with the symmetry requirements of the viscous force (the force must flip sign under flow reversal) and that (ii) respects the rotational invariance of the system under global rotations is
\begin{align}
	G_b[\rho,\vel,\acc]&=\int d\rv\int_0^t dt'K_b^{\vel}(t-t')\rho'(\nabla\cdot\vel')(\nabla\cdot\acc)\rho\nonumber\\
	&+\int d\rv\int_0^tdt'K_b^{\acc}(t-t')\rho'(\nabla\cdot\acc')(\nabla\cdot\acc)\rho,\label{eq:Gb}\\
	G_s[\rho,\vel,\acc]&=\int d\rv\int_0^t dt'K_s^{\vel}(t-t')\rho'(\nabla\times\vel')\cdot(\nabla\times\acc)\rho\nonumber\\
	&+\int d\rv\int_0^tdt'K_s^{\acc}(t-t')\rho'(\nabla\times\acc')\cdot(\nabla\times\acc)\rho,\label{eq:Gs}
\end{align}
where we have omitted the dependencies of the one-body fields, e.g.\ $\rho=\rho(\rv,t)$, primed fields are evaluated at $t'$, e.g. $\rho'=\rho(\rv,t')$, and the spatial integral runs over the whole system.
Analogue expressions arise in overdamped Brownian dynamics based on an expansion in terms of the velocity gradient $\nabla\vel$~\cite{PRLnablaV}.

The shear $\fs$ and bulk $\fb$ viscous forces shown in Eqs. (1) and (2) of the main text are then generated via the functional derivative
\begin{equation}
	\textbf{f}_\alpha(\rv,t)=-\frac{\delta G_\alpha}{\delta\Jdot(\rv,t)}=-\frac1{\rho}\frac{\delta G_\alpha}{\delta\acc(\rv,t)},\;\;\;\;\;\alpha=b,s,\label{eq:derivative}
\end{equation}
where the derivative is taken at time $t$ with respect to either $\Jdot$ or $\acc$ and considering that the fields $\rho(\rv,t'),\vel(\rv,t'),$ and $\acc(\rv,t')$ are kept fixed at their real physical values for all previous times $t'<t$.
Hence, the functional derivative in Eq.~\eqref{eq:derivative} acts only on the unprimed terms $\acc(\rv,t)$ of Eqs.~\eqref{eq:Gb} and~\eqref{eq:Gs} but not on the primed $\acc'=\acc(\rv,t')$ ones.

\subsection{Shear flow in Brownian dynamics}\label{8}
\begin{figure}
    \centering
    \resizebox{0.5\textwidth}{!}{\includegraphics{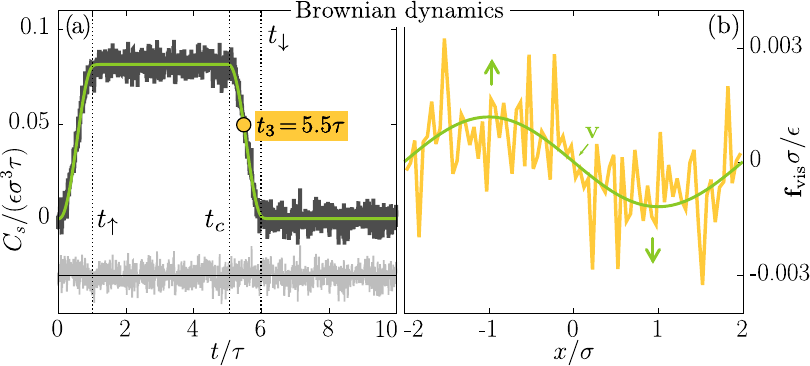}}
	\caption{
	(a) Temporal dependency of the shear viscous force $C_s$ as a function of time $t$ in Brownian dynamics simulations (thick black line)
        and theoretically (green) for the shear flow described in the main text.
        The vertical dotted lines indicate the times $t_\uparrow, t_c$, and $t_\downarrow$.
        The time $t_3=5.5\tau$ is highlighted with a yellow circle.
        The light grey line fluctuating around the horizontal line is the difference between simulation (thick black) and theory (violet).
        (b) Shear viscous force $\fvis$ as a function of $x$ at time $t_3=5.5\tau$ according to BD (yellow) and theory (green).
        The force points along the $y$-axis.
        The colored arrows indicate the direction of the force at the selected positions.
	    }
    \label{fig:BDshear}
\end{figure}
As in the case of the compressible flow, we have also analysed the shear flow using Brownian dynamics simulations.
Since the acceleration field does not play any role in overdamped Brownian dynamics, the velocity field alone reproduces the complete shear viscous force.
We show in Supplementary Fig.~\ref{fig:BDshear} the temporal part $C_s(t)$ of the viscous force vs time, and the viscous force vs the $x$-coordinate for a given time obtained in Brownian dynamics simulations along with the corresponding theoretical predictions [kernel parameters $\cvs/(\epsilon \sigma^3 \tau)=0.081$ and $\tvs/\tau=0.059$].
The parameters of the flow are identical to those used in MD (see main text).

\subsection{Traveling shear wave}\label{9}
\begin{figure*}
    \centering
    \resizebox{0.9\textwidth}{!}{\includegraphics{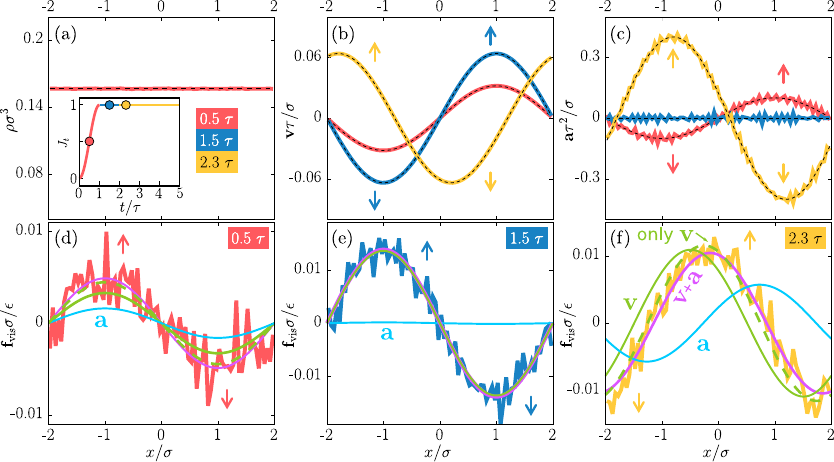}} 
	\caption{Density (a), velocity (b), and acceleration (c) profiles vs the $x$-coordinate for the shear flow with a traveling wave.
	Three different times $0.5\,\tau$ (red), $1.5\,\tau$ (blue), and $2.3\,\tau$ (yellow) are shown.
	Solid lines are simulation data sampled using custom flow and thin black-dashed lines are the corresponding target fields.
	The inset in (a) shows the temporal dependency of the current with the three different regimes highlighted using different color and the selected times indicated by colored circles.
	The amplitude of the current increases until $t/\tau=1$ (red) and then it remains constant until the simulation ends. The shear wave is stationary at first (blue) and then it starts to travel for $t/\tau>2$ (yellow).
	Panels (d),(e), and (f) show the shear viscous force at different times (as indicated) obtained in simulations (thick solid lines) along with the theoretical prediction using the acceleration and the velocity contributions (violet) and also using only the velocity contribution (dashed green). The individual contributions of the velocity (solid green) and of the acceleration (blue) to the total signal (violet) are also shown. Once the shear wave is traveling (f) there is a clear phase shift between the simulation data (yellow) and the prediction using only the velocity (dashed green). Using both contributions (violet) the simulation data is correctly reproduced. The color arrows indicate the direction of the respective vector field at the selected positions.
	}
    \label{fig:shearwave}
\end{figure*}
For the traveling shear wave, the current follows up to $t=2\tau$ the same time evolution as in the shear case, see Eqs. (8) and (9) of the main text.
After $t=2\tau$, the shear wave starts to move with constant velocity $v_s= 4 \sigma/\tau$ and constant amplitude, i.e.\
\begin{equation}
	\J(\rv,t)=J_0\sin\left(\dfrac{2\pi (x-v_st)}{L_x}\right)\ey,\;\;\; t>2\tau,
\end{equation}
where, as in the non-traveling shear case, $L_x/\sigma=4$ and $J_0\tau\sigma^2=0.01$.
Representative states for each of the regimes of the traveling shear flow are shown in Supplementary Fig.~\ref{fig:shearwave}.
The density (a) remains constant in space at every time. The amplitudes of the velocity (b) and of the acceleration (c) increase until $t=1\tau$.
To relax memory effects, the velocity profile remains stationary from $1\tau$ to $2\tau$.
Hence, the acceleration vanishes everywhere in that time period.
Then, the traveling wave begins to move and the velocity field changes its phase with constant speed.
Therefore, the acceleration field has a constant instantaneous phase shift of $\pi/2$ with respect to the velocity field.
This is different from what we considered in the static shear wave and allows us to test our model for the shear viscous force.
The shear viscous forces at three different times are shown in panels (d),(e), and (f) of Supplementary Fig.~\ref{fig:shearwave}.
We show the data sampled in molecular dynamics simulations along with the theoretical predictions which we calculate with the same kernel parameters previously obtained for the static shear wave flow.
The agreement between simulation and theory is excellent, not only before the wave starts to travel (d,e), which was expected from the static shear case, but also during the traveling wave (f).
The phase shift between $\vel$ and $\acc$ has an effect on the viscous force that is theoretically reproduced. 
In contrast, if we use only the velocity dependent part of the viscous force and the same kernel parameters as for the static case, there is a phase shift between the theoretical predictions and the simulation data.
The acceleration field is therefore required to describe the data accurately.